\documentclass{article}
\usepackage[a4paper,margin=1in,footskip=0.25in]{geometry}
\usepackage{graphicx}
\usepackage{subfigure}
\usepackage{color}
\usepackage{esvect}
\usepackage{cancel}
\usepackage{physics}
\usepackage{soul}
\usepackage{lipsum}
\usepackage{amsfonts}
\usepackage{esvect}
\usepackage{physics}
\usepackage{url}

\newcommand{\Var}{\mathrm{Var}}

\newcommand{\pinfrac}{\frac{n_{r}p_{\text{in}}}{n_{r}p_{\text{in}}+(N-n_r) p_{\text{out}}}}

\usepackage[a4paper,margin=1in,footskip=0.25in]{geometry}
\usepackage{subfigure}
\usepackage[subfigure]{tocloft}
\usepackage{setspace}
\usepackage{graphicx}
\usepackage{algorithm}
\usepackage[noend]{algpseudocode}
\usepackage{color}
\usepackage{amssymb,amsfonts,amsmath,amsthm}

\title{Bias and variance in the social structure of gender}
\author{  
Kristen M. Altenburger
\thanks{
Department of Management Science \& Engineering, Stanford University.
Email: \texttt{kaltenb@stanford.edu}.
}
 \and 
Johan Ugander
\thanks{
Department of Management Science \& Engineering, Stanford University.
Email: \texttt{jugander@stanford.edu}.
}
}
\date{}

\begin{document}
\maketitle

\begin{abstract}
The observation that individuals tend to be friends with people who are similar to themselves, commonly known as homophily, is a prominent and well-studied feature of social networks. Many machine learning methods exploit homophily to predict attributes of individuals based on the attributes of their friends. Meanwhile, recent work has shown that gender homophily can be weak or nonexistent in practice, making gender prediction particularly challenging. In this work, we identify another useful structural feature for predicting gender, an overdispersion of gender preferences introduced by individuals who have extreme preferences for a particular gender, regardless of their own gender. We call this property monophily for ``love of one,'' and jointly characterize the statistical structure of homophily and monophily in social networks in terms of preference bias and preference variance. For prediction, we find that this pattern of extreme gender preferences introduces friend-of-friend correlations, where individuals are similar to their friends-of-friends without necessarily being similar to their friends. We analyze a population of online friendship networks in U.S.~colleges and offline friendship networks in U.S.~high schools and observe a fundamental difference between the success of prediction methods based on friends, ``the company you keep,'' compared to methods based on friends-of-friends, ``the company you're kept in.'' These findings offer an alternative perspective on attribute prediction in general and gender in particular, complicating the already difficult task of protecting attribute privacy.
\end{abstract}

\bigskip
Homophily is the observed phenomenon in social networks whereby friendships form frequently among similar individuals \cite{lazarsfeld1954friendship,mcpherson2001birds}. 
Homophily can originate from an individual's personal preference to become friends with similar others (choice homophily), structural opportunities to interact with similar others (induced homophily), or a combination of both \cite{kossinets2009origins}. 
An important consequence of homophily is that even if an individual does not disclose attribute information about themselves (such as their gender, age, or race), methods for relational learning \cite{neville2002supporting,jensen2004collective,macskassy2007classification,sen2008collective,bhagat2011node,taskar2002discriminative} can often leverage attributes disclosed by that individual's friends to predict their private attributes. Gender prediction, however, is a difficult relational learning problem, as gender homophily can be weak or non-existent in both online and offline settings \cite{ugander2011anatomy,thelwall2009homophily,shrum1988friendship,neal2010hanging,laniado2016gender}. 
Weak gender homophily motivates us to examine alternative network structures useful for attribute prediction \cite{duncan1986disclosure}.

In this work, we focus on gender prediction and document the presence of individuals in social networks with extreme gender preferences for a particular gender, regardless of their own gender. We call this overdispersion of preferences ``monophily'' to indicate it as distinct from the preference bias introduced by homophily, and observe that monophily is nearly ubiquitous across the population of online and offline friendship networks that we study. The presence of these individuals with extreme preferences introduces similarity among friends-of-friends or along 2-hop relations. For the practical problem of attribute prediction, being friends with an individual with extreme gender preferences is a strong signal of one's own gender and is therefore useful for gender prediction.

In order to model these empirical observations, as part of this work we also introduce an overdispersed stochastic block model that enables us to separately simulate homophily and monophily in social networks. We show how the 2-hop structural relationship induced by overdispersion (monophily) can exist in the complete absence of any 1-hop bias (homophily), and find that overdispersed friendship preferences can drive successful classification algorithms in settings with weak or even no homophily. Therefore, in networks with weak homophily but strong monophily, your friends-of-friends (``the company you're kept in'') can then be responsible for disclosing private attribute information, as opposed to your friends (``the company you keep''). These findings extend the importance of privacy policies that protect relational data, while also proposing an intuitive structural property of social networks of independent interest.

In the spirit of a solution-oriented science \cite{watts2017should}, our analysis addresses the practical problem of inferring gender on social networks by revisiting the social theory of homophily and introducing alternative considerations for heterogeneity in friendship preferences. In addition to improving prediction, we also present monophily as an independent structure of interest when studying ``gender as a social structure'' \cite{risman2004gender} by explicitly quantifying the variability in gender preferences beyond the bias captured by homophily.  Only recently has the role of variability in general and overdispersion in particular been studied on social networks where classic perspectives have prioritized analyzing aggregate patterns of interaction \cite{raftery2001statistics}. This work follows other advances in incorporating variance and overdispersion in social data analysis include understanding the consequences of overdispersion when estimating the size of sub-populations \cite{zheng2006many}, documenting variations in the homophily of political ideology \cite{boutyline2016social}, assessing gender variation in linguistic patterns \cite{bamman2014gender}, and inferring social structure based on indirectly observed data \cite{mccormick2013practical}. 

The paper proceeds by first establishing how we measure the bias (homophily) and excess variance (monophily) of gender preferences. We then examine how relational inference methods for node classification relate to the presence of homophily and/or monophily. While previous models of homophily have shown its statistical significance in network data \cite{wimmer2010beyond,smith2014social}, we highlight that the statistical significance of homophily does not necessarily imply predictive power when the task is to infer private attributes. Following the empirical analysis, we introduce a network model of overdispersed preferences that generalizes the well-studied stochastic block model \cite{holland1983stochastic}. Throughout this work we view gender as a binary attribute and aim to measure homophily in a manner that encompasses all sources of preference due to both choice and induced homophily. While we focus on gender, the methods developed in this work contribute a broad statistical toolkit for the general study of variability in social group interactions across a wide range of attributes or traits. 

We begin by showing how the conventional homophily index can be interpreted as the maximum likelihood estimate of a parameter within a simple generalized linear model. We then extend this model to capture overdispersed preferences using a quasi-likelihood approach, introducing an overdispersed model with additional parameters that concisely measure the overdispersion of gender preferences among females ($F$) and males ($M$), respectively. We propose estimates of these parameters as our measures of monophily among females and males in network data.

The homophily index of a graph \cite{coleman1958relational,currarini2009economic} characterizes the aggregate pattern of individuals' biases or preferences in forming friendships with people of their own attribute class relative to people from other classes. For a generic attribute class $r$ and assuming there are $k = 2$ classes, 
the homophily index with respect to class $r$ is defined as
\begin{eqnarray}
\label{eq:hr}
\hat{h}_r=\frac{ \sum_{i\in r}d_{i,\text{in}} }{ \sum_{i\in r} d_{i,\text{in}} + \sum_{i\in r} d_{i,\text{out}}}=\frac{ \sum_{i\in r}d_{i,\text{in}} }{ \sum_{i\in r} d_i },
\end{eqnarray}
where $d_{i,\text{in}}$ denotes node $i$'s observed $in$-class degree with similar others, $d_{i,\text{out}}$ denotes its observed $out$-class degree with different others, $d_{i}$ denotes its observed total degree, and $n_r$ will represent the total number of nodes with attribute $r$ such that $N = \sum_{r=1}^k n_r$. For notational simplicity, we use $i\in r$ to refer to the set of all nodes with attribute value $r$.

In measuring binary gender homophily (i.e.~$r=F$ or $r=M$), we first illustrate how to measure homophily among females. We assume that each individual $i \in F$ in a network forms $in$-class connections with the other $n_F$ individuals at a rate $p_{\text{in,F}}$ and $out$-class ties with the other $n_M$ individuals at a rate $p_{\text{out}}$ (and similarly for each individual $i \in M$ that a connection with males form at a rate $p_{\text{in,M}}$ and with females form at a rate $p_{\text{out}}$). We therefore expect for each individual $i \in F$ that their class-specific degrees obey the following distributions (permitting self-loops):
\begin{eqnarray}
D_{i,{\text{in}}} | p_{\text{in,F}}  \sim & \text{Binom}(n_F, p_{\text{in,F}}), \label{eq:2}\\ 
D_{i,{\text{out}}} | p_{\text{out}} \sim & \text{Binom}(n_M, p_{\text{out}}), \label{eq:3}\\
D_i|p_{\text{in,F}},p_{\text{out}} = & D_{i,{\text{in}}}| p_{\text{in,F}} + D_{i,{\text{out}}}| p_{\text{out}},
\label{eq:4}
\end{eqnarray} 
where $D_{i,\text{in}}$ is a random variable describing the $in$-class degree, $D_{i,\text{out}}$ describes the $out$-class degree, and $D_i$ describes the total degree of node $i$ in class $F$. We explicitly condition these random variables on the parameters $p_{\text{in,F}}$ and $p_{\text{out}}$ to make clear that these parameters are, for now, fixed and constant. 

The nodes $i \in M$ have the same binomial degree distribution specified by $in$- class degrees formed among the $n_M$ nodes at a rate $p_{\text{in,M}}$ and $out$- class degrees formed among the $n_F$ nodes at a rate $p_{\text{out}}$. With only $k=2$ classes, for simplicity we use the notation $p_{\text{out}}$ in place of e.g.~$p_{\text{out,r,s}}$, highlighting that the rates could depend on the specific $in$- and $out$-classes $r$ and $s$ in the most general directed multi-class case. Note that the random variables in equations~(\ref{eq:2})--(\ref{eq:4}) are approximately independent, but not completely: constraints on the joint distribution of the degrees corresponding to the constraints of the Erd\H{o}s-Gallai theorem (since the degrees must correspond to a graph) create a dependence, but this dependence is small for graphs of modest size or larger \cite{van2016random} and we safely ignore it here. 

To show how the homophily index can be estimated using a generalized linear model (GLM) \cite{mccullagh1989generalized} of $in$- versus $out$-class degrees, let the observed degree data be $\{(d_{i,\text{in}}, d_i), i \in F\}$, where the set-up is analogous for $i \in M$. Among the $F$ individuals, their $in$-class degree distribution conditional on their total observed degree is approximately distributed as 
\begin{eqnarray}
D_{i,\text{in}} | d_i, p_{\text{in,F}}, p_{\text{out}} \sim \text{Binom}(d_i,n_{F}p_{\text{in,F}}/(n_{F}p_{\text{in,F}}+n_{M}p_{\text{out}}))
\end{eqnarray}
in the case of two attribute classes (Supplementary Note 1). By applying a logistic-binomial model \cite{gelman2006data,agresti2011categorical}, an adaptation of the logistic regression model for count data, the logistic link function of the binomial logistic regression model is then specified as $n_{F}p_{\text{in,F}}/(n_{F}p_{\text{in,F}}+n_{M}p_{\text{out}}) = \text{logit}^{-1}(\beta_{0F}) = e^{\beta_{0F}}/(1+e^{\beta_{0F}})$ assuming there are no additional covariates 
(which could otherwise be incorporated). 
For this model we can then derive the maximum likelihood estimate of $\beta_{0F}$ as: 
\begin{eqnarray}
\hat{\beta}^{MLE}_{0F} = \text{logit}(\sum_{i\in F} d_{i,\text{in}}/\sum_{i\in F} d_{i} )= \text{logit}( \hat{h}_F),
\end{eqnarray}
or equivalently $e^{\hat{\beta}^{MLE}_{0F}}/(1+e^{\hat{\beta}^{MLE}_{0F}}) 
= \hat{h}_F$ (Supplementary Note 2). Here $\hat{h}_F$ is exactly the homophily index specified in equation \eqref{eq:hr} above, and hence the homophily index can be interpreted as the intercept term estimated from a GLM applied to the observed degree data.

Given this interpretation of the homophily index within a GLM framework, it is useful to refer to the quantity $n_{r}p_{\text{in,r}}/(n_{r}p_{\text{in,r}}+(N-n_r) p_{\text{out}}) = h_r = \text{logit}^{-1}(\beta_{0r})$ as the ``homophily parameter'' for each class $r$, letting the ``homophily index'' for each class embody the corresponding maximum likelihood estimate, $\hat{h}_r$. The homophily index is focused on assessing whether $\hat{h}_r$ is different from the $in$-class' relative proportion in the population, $n_{r}/N$. Meanwhile, this model gives a poor assessment of the variance of the data due to the constrained relationship between mean and variance \cite{gelman2006data}. More specifically, in this model of $in$-class degrees for class $r$, the variance of the $in$-class degrees is constrained to be $\Var [D_{i,\text{in}} | d_i ] =  d_i h_r  \left(1 - h_r\right)$ (Supplementary Note 3). 

\begin{figure}
\begin{center}
{
	\includegraphics[width=0.8\textwidth]{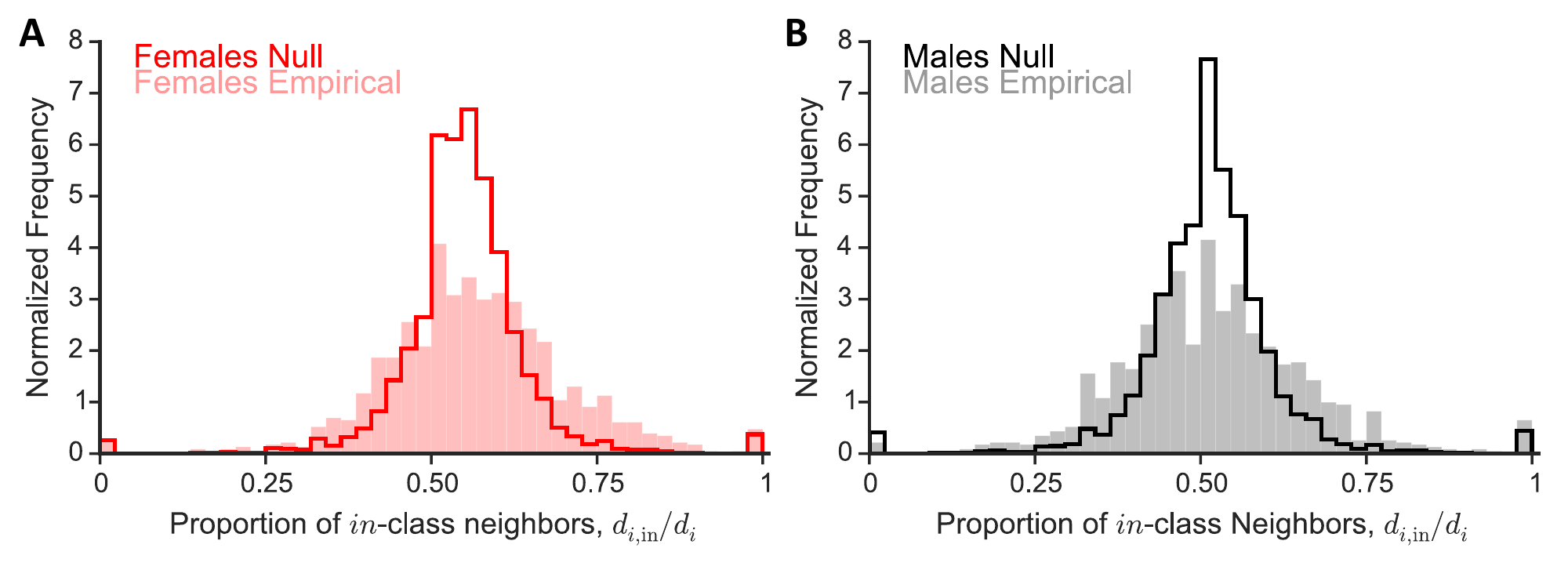}
\caption{
Evidence of overdispersion in gender preferences. On the Amherst College network we compute the empirical distribution (filled bars) of $in$-class preferences for females (Left) and males (Right).
We compare these distributions to a null distribution (solid lines) based on preferences with binomial variation (for details of null model sampling, see Methods). We observe overdispersion of $in$-class gender bias in friendship formation for females and males as the observed empirical variance is greater than under the null.}
\label{fig:fb_hom1}
}
\end{center}
\end{figure}

We observe that across the full population of 97 co-educational college online social networks from the Facebook100 dataset (FB100), the distribution of gender preferences are overdispersed, with a variance larger than the above model predicts (for details on the FB100 dataset, see Methods). 
As seen in Figure~\ref{fig:fb_hom1} for the Amherst College network, the empirical distributions of the gender preferences are more dispersed (less concentrated) than the homophily-only null distributions (for details of null model sampling, see Methods). Across the females and males at Amherst College, there is clear evidence that the variance of the distribution of $in$-class preferences is greater than what would be expected given the homophily-only null model.

We formally test the statistical significance of overdispersion of $in$-degrees relative to $out$-degrees among nodes with attribute class value $r$ given the fitted GLM with $\hat{\beta}^{MLE}_{0r} = \text{logit}(\hat{h}_r)$ and the nominal variance of individual $i$'s $in$-class degree count under this model. The standard test for overdispersion compares the sum of squared standardized residuals $\sum_{i\in r} \frac{(d_{i,\text{in}} - d_i \hat{h}_r)^2}{d_i \hat{h}_r  (1-\hat{h}_r)}$ to $\chi^2_{n_r - 1}$, where there are $n_r - 1$ degrees of freedom since the model features only a single intercept parameter \cite{williams1982extra,gelman2006data} (Supplementary Note 4). We consistently observe the variance of $in$-class degrees among the females and males are significantly greater than what can be explained by the homophily-only GLM across the 97 college networks, with $p<10^{-3}$ for all networks. The friendship networks in the Add Health dataset show equivalent evidence of overdispersion in a directed setting (for details on the Add Health dataset, see Methods). 

A variety of modeling methods have been proposed to measure and model extra variation in count data \cite{wedderburn1974quasi,williams1982extra,mccullagh1989generalized,morel1993finite}. We employ a quasi-likelihood approach \cite{williams1982extra}, the least presumptive approach compared to alternative methods, in order to adapt the GLM to accommodate this overdispersion. The quasi-likelihood set-up allows each node $i$ in class $r$ to have an individual latent preference for $in$-class friendships, $h_{i,r}$, such that $\mathbb{E}[h_{i,r}]=h_r$ and $\Var [ h_{i,r} ] = \phi_r  h_{r} (1 - h_{r})$ for some $\phi_r \ge 0$. The parameter $\phi_r$ is introduced to incorporate the extra variation, and the variance is parameterized as such for notational convenience (Supplementary Note 3). This set-up does not specify a distribution on $h_{i,r}$ but instead uses $\phi_r$ to quantify how much nodes in class $r$ vary in allocating their $in$-class versus $out$-class friendships.

The case when $\phi_r=0$ corresponds to the typical homophily-only model (Williams' Model I), which restricts $h_{i,r}$ to be constant across all nodes in the class. Letting $\phi_r>0$ (Williams' Model II) captures variation beyond the conventional model (Supplementary Note 3). Through an iterative procedure due to Williams that maximizes a quasi-likelihood function (Supplementary Note 4), we jointly estimate $\hat{\beta}_{0F}^{MQE}, \hat{\phi}_F$ among female nodes and $\hat{\beta}_{0M}^{MQE}, \hat{\phi}_M$ among male nodes, allowing us to use $\hat{\phi}_F$ and $\hat{\phi}_M$ as measures of preference overdispersion in the data. Note that the homophily measures estimated under Williams' Model II, $\hat{\beta}_{0F}^{MQE}$ and $\hat{\beta}_{0M}^{MQE}$, are slightly different than the traditional homophily indices, $\hat{\beta}_{0F}^{MLE}$ and $\hat{\beta}_{0M}^{MLE}$, but the estimates $\hat{\beta}_{0r}^{MLE}$ and 
$\hat{\beta}_{0r}^{MQE}$ are highly correlated (Supplementary Note 5), and we focus our characterization of homophily on $\hat{\beta}^{MLE}_{0F}$ and $\hat{\beta}^{MLE}_{0M}$ given the direct connection to the homophily index.

\begin{figure}
\begin{center}
{
	\includegraphics[width=0.8\textwidth]{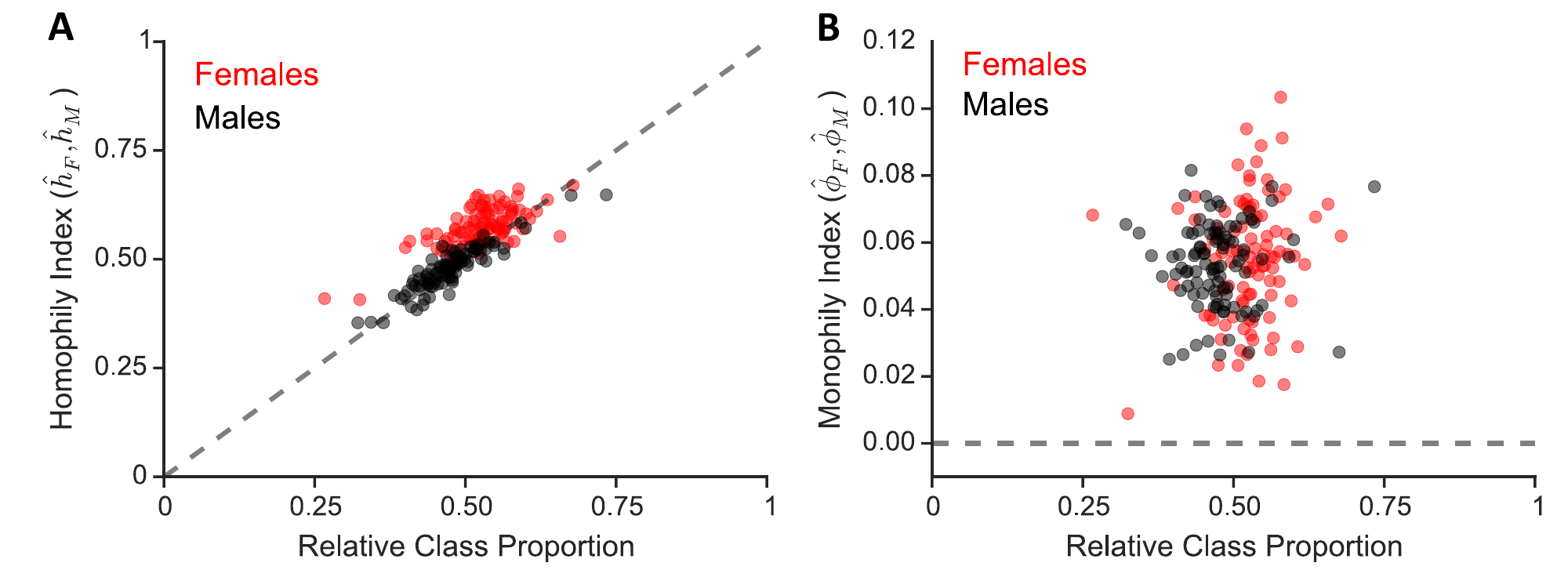}

\caption{
Homophily and monophily across the population of friendship networks. Measuring homophily and monophily in social networks. (Left) The homophily index $\hat{h}_r$ and (Right) monophily index $\hat{\phi}_r$ for bias and overdispersion, respectively, in friendship formation among male and female students at each of 97 online college social networks. The homophily indices are concentrated around relative class proportions (dashed line), while the monophily indices all show overdispersed preferences independent of the relative class proportions. Dashed lines indicate the lines of no homophily and no monophily, respectively.
}
\label{fig:fb_hom2}
}
\end{center}
\end{figure}

In Figure~\ref{fig:fb_hom2}, we evaluate both bias (homophily) and overdispersion (monophily) in gender preferences, using the conventional homophily index $\hat{h}_r$ to measure bias and the estimates $\hat{\phi}_r$ to measure overdispersion across the populations of college networks in the FB100 dataset. 
We see that across these networks the homophily measures $\hat{h}_r$ closely follow the class proportion $n_r/N$, whereas the monophily measures $\hat{\phi}_r$ depart significantly from zero and show no sign of varying with class proportion. We next show how overdispersed preferences help explain the ``predictability'' of gender in relational trait inference in settings with weak or nonexistent gender homophily.

Having established $\hat{\phi}_F$ and $\hat{\phi}_M$ as our measures of overdispersion, we now illustrate the key role overdispersion can play in the success of some but not all methods for relational inference. Our specific focus is to understand how the efficacy of different relational inference methods varies in the presence or absence of homophily and/or monophily, building on the challenge of predicting gender on large-scale social networks with minimal gender homophily. We explore a typical setting where individuals reveal information completely at random \cite{he2006inferring,macskassy2007classification,rubin1976inference,heitjan1996distinguishing} (i.e.~uniformly), meaning that the likelihood to be labeled or to provide public information does not depend on other attributes. The prediction task is then to infer private gender attributes using public gender attributes and the social network relationships.  We address this prediction problem through the lens of homophily and monophily. While historically the social sciences have placed a strong emphasis on explanation at the expense of prediction \cite{hofman2017prediction}, this work reverses this traditional focus by showing how statistically significant homophily does not necessarily imply high predictability of attributes. Instead, we highlight the role of variation in relational inference methods, especially in applications when the bias introduced by homophily is weak or nonexistent.

Relational inference methods can be categorized based on the neighborhood relationships they exploit for classification, either learning from 1-hop (immediate friends) or 2-hop (friend-of-friend) relations. This distinction in relational learning is not often considered, but we note that it is a direct analog of a common distinction between the PageRank \cite{page1998pagerank} and Hubs and Authorities \cite{kleinberg1998authoritative} algorithms in graph ranking. PageRank is based on the principle that ``a node is important if it is linked to by other important nodes,'' while Hubs and Authorities is based on the principle that ``a node is important if it is linked to by nodes that link to important nodes.'' These differing principles can extract very different notions of importance in graph ranking; the latter is motivated by web ranking problems where, e.g., car companies don't link to other car companies but should still appear high in search results for ``cars.'' Analogously, we observe that 2-hop and 1-hop methods are differently well-suited for different node classification problems. We compare these classification methods relative to a baseline model that assigns scores based on the relative class proportions observed in the training sample.

Classification methods based on a node's 1-hop (immediate) neighbors include:
\begin{itemize}
\item The 1-hop Majority Vote (1-hop MV) classifier, also called the weighted-vote relational neighbor (wvRN) classifier \cite{macskassy2007classification}, builds directly on similarities between connected nodes where unlabeled nodes are scored based on the proportion of labels among their neighbors. When a node does not have any labeled neighbors, the relative class proportions in the training data are used (Supplementary Note 6). 

\item The ZGL method \cite{zhu2003semi} scores unlabeled nodes by computing the relative probabilities of reaching each node in a graph under a random walk originating at the labeled node sets. The ZGL method can be characterized as an iterated/semi-supervised adaptation of 1-hop MV \cite{bhagat2011node}.

\end{itemize}
Methods that exploit 2-hop (neighbor-of-neighbor) relations include:
\begin{itemize}
\item The 2-hop Majority Vote (2-hop MV) classifier uses the relationship between a node and its 2-hop neighbors weighted by the number of length-2 paths. Unlabeled nodes are scored based on the weighted proportion of labels among their 2-hop neighbors.

\item LINK-Logistic Regression \cite{zheleva2009join} uses labeled nodes to fit a regularized logistic regression model (Supplementary Note 7) that interprets rows of the adjacency matrix as sparse binary feature vectors, striving to predict labels from these features. The trained model is then applied to the feature vectors (adjacency matrix rows) of unlabeled nodes, which are scored based on the probability estimates from the model. Small variations that use the same feature set but employ e.g.~SVMs or Random Forests instead of Logistic Regression give qualitatively similar performance. Employing the LINK feature set as part of a Naive Bayes classifier gives a clear view of LINK as a family of 2-hop methods (Supplementary Note 8).

\end{itemize}

\begin{figure}
\begin{center}
{
	\includegraphics[width=0.6\textwidth]{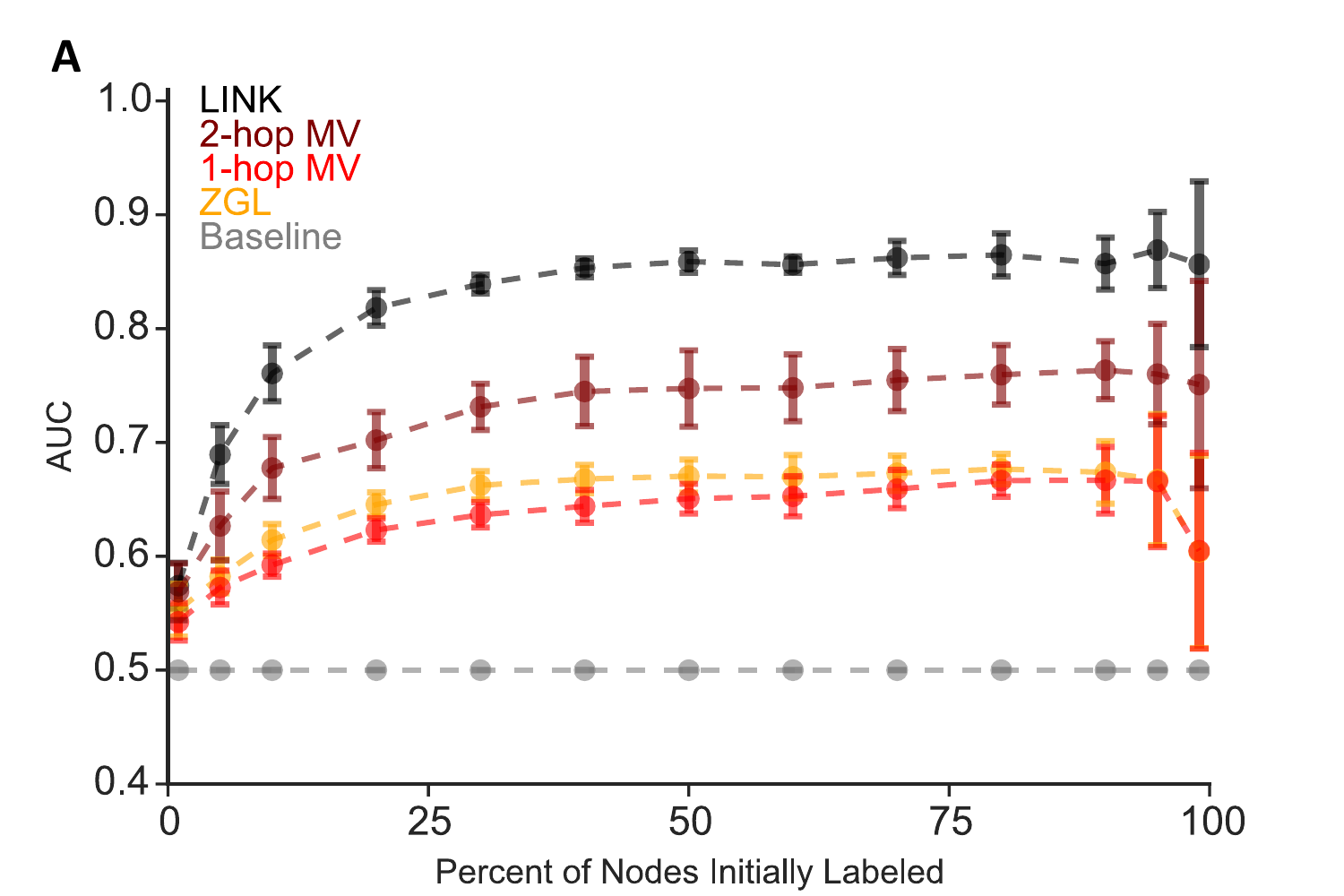}
	\includegraphics[width=0.6\textwidth]{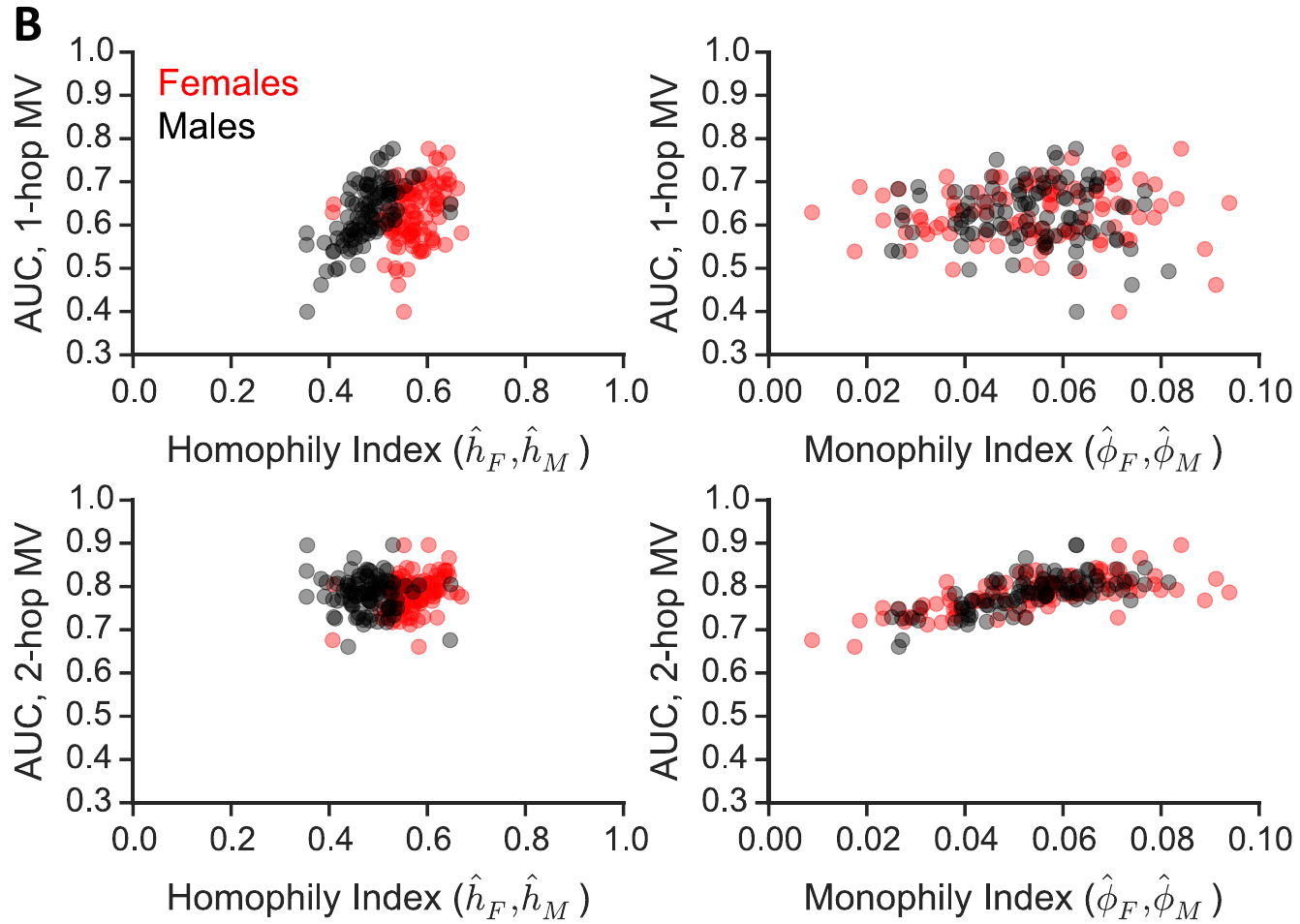}
}
\end{center}
\caption{
Comparison of 1-hop versus 2-hop classifiers and the relationship between classification performance and homophily versus monophily. (Top) A comparison of the performance of classification methods for gender inference on the Amherst College network with $n_F$=1015 and $n_M$=1017, measured by AUC, varying the percentage of nodes that are given as labeled (for details on the cross-validation, see Methods). Homophily and monophily measured for the Amherst College give $\hat{h}_F=0.55$, $\hat{\phi}_F=0.04$ and $\hat{h}_M=0.51$, $\hat{\phi}_M=0.04$. We observe strong classification performance from the LINK method, which we attribute to the overdispersed gender preferences. (Bottom) Across FB100 networks we compare the correlation between 1-hop and 2-hop Majority Vote (with 50\% initially labeled nodes) versus gender homophily and gender monophily. We observe that homophily has high explanatory power for the 1-hop Majority Vote AUC across schools while monophily has very little. Meanwhile, homophily has weak explanatory power of the 2-hop Majority Vote AUC across schools while monophily has strong explanatory power for that method.}
\label{fig:inference}
\end{figure}

We observe only slight gender homophily across the population of college networks in the FB100 dataset, and accordingly in Figure~\ref{fig:inference}A we observe limited performance using 1-hop methods (1-hop MV and ZGL) to predict gender in a single representative network. Meanwhile, we see that 2-hop methods (2-hop MV and LINK) have higher performance, corroborating our intuition for 2-hop methods being able to surface structural signals for classification in the presence of overdispersed preferences. As illustrated in Figure \ref{fig:inference}B, classification for 2-hop Majority Vote considerably outperforms classification based on 1-hop Majority Vote across the population of FB100 schools, and we attribute this performance difference to the monophily in the network. In addition to the undirected FB100 networks, we also examined node classification on the directed Add Health school networks (Supplementary Note 9), where we observe similar results. 

In order to generalize these empirical observations on the impact of homophily versus monophily on 1-hop and 2-hop inference methods, we generate synthetic graphs with extra-binomial variation by introducing a variant on the stochastic block model (SBM) \cite{holland1983stochastic}, also known as the planted partition model \cite{condon2001algorithms}, a well-studied statistical distribution over graphs with desired block structure commonly employed to study network association patterns. An SBM models association preferences among $k$ node classes by specifying a set of block sizes $n_1, \ldots, n_k$ and a preference matrix $\mathbf{P}$ where $\mathbf{P}_{{a_i}{a_j}}$ denotes the independent probability of an edge between nodes $i$ and $j$ in attribute classes $a_i$ and $a_j$. For modeling associations between two genders using SBMs, the matrix $\textbf{P}$ is simply a $2 \times 2$ matrix denoting the edge probabilities within and between the two genders. Assortative block structure is present when $in$-class probabilities are greater than $out$-class probabilities.

We propose an overdispersed extension of the stochastic block model to additionally capture monophily (extra-binomial heterogeneity in preferences) by relaxing this restriction of fixed class probabilities among all nodes in a given class and assuming a latent distribution on gender preferences \cite{williams1982extra}. We specifically employ a latent Beta distributions on preferences \cite{crowder1978beta} applied to graphs, though other latent distributions or other means of incorporating overdispersion \cite{diprete1994multilevel, guo2000multilevel} could be just as reasonable; note that the measure of monophily $\hat{\phi}_r$ developed earlier in this work (that uses a quasi-likelihood approach) is agnostic to the choice of latent distribution.

The proposed overdispersed stochastic block model (oSBM) is defined by the block sizes $n_1,\ldots,n_k$, $k\times k$ preference matrix $\mathbf{P}$, and additional overdispersion parameters $\phi_{\text{in}}^* \ge 0$ and $\phi_{\text{out}}^* \ge 0$. 
Here $\phi_{\text{in}}^*$ and $\phi_{\text{out}}^*$ are concrete parameters of a generative model, while we will continue to use 
$\phi_r$ to describe generic overdispersion in preferences (when $\phi_r>0$). Networks are generated from the model via a multi-level approach, where first each node's $in$- and $out$-class degrees are created by sampling class preference parameters ($p_{i,\text{in}}$ and $p_{i,\text{out}}$) from an appropriate latent Beta distribution with specified means $p_{\text{in}}$ and $p_{\text{out}}$ for $in$- and $out$- class probabilities respectively. We assume the same mean across all attribute classes $r$, so we denote this mean by $p_{\text{in}}$ instead of $p_{\text{in,r}}$ for a given class $r$. Given the resulting individual preferences, a graph is generated analogously to how the degree-corrected SBM \cite{karrer2011stochastic} attains prescribed degrees using a Chung-Lu construction \cite{chung2002connected}, with expected $in$-degrees $d_{i,\text{in}} = n_r p_{i,\text{in}}$ and expected $out$-degrees $d_{i,\text{out}} = (N-n_r) p_{i,\text{out}}$ (Supplementary Note 10). We note that this overdispersed stochastic block model complements related work on overdispersion in social network surveys \cite{zheng2006many} where an individual's degree to a class is taken to be distributed Gamma-Poisson. Under an oSBM, the number of individuals from a specific class that a given node is connected to will approximately follow a Beta-Binomial distribution, a close relative of the Gamma-Poisson distribution \cite{chatfield1976beta}.

\begin{figure}
\begin{center}
{
	\includegraphics[width=0.75\textwidth]{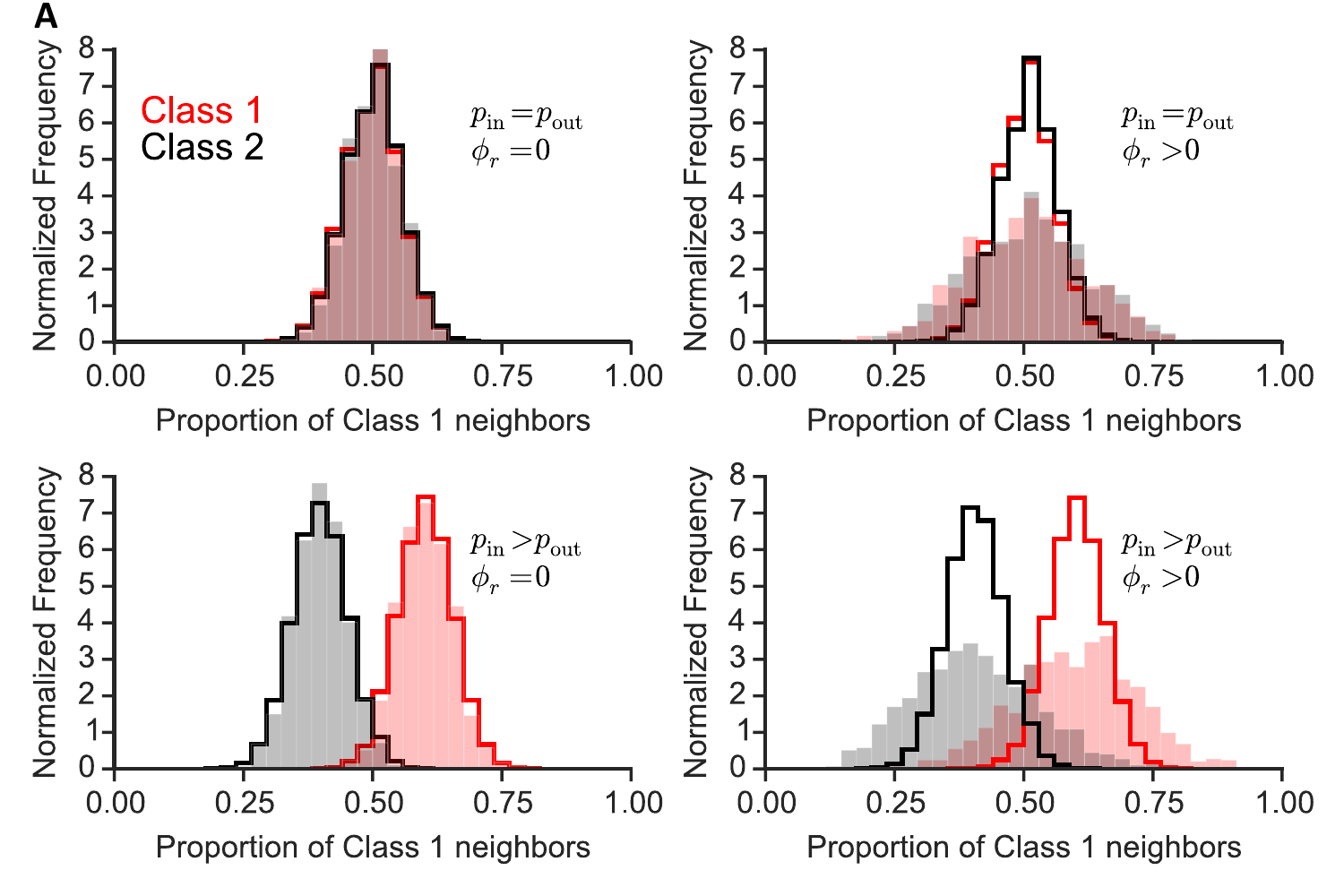}
	\includegraphics[width=0.75\textwidth]{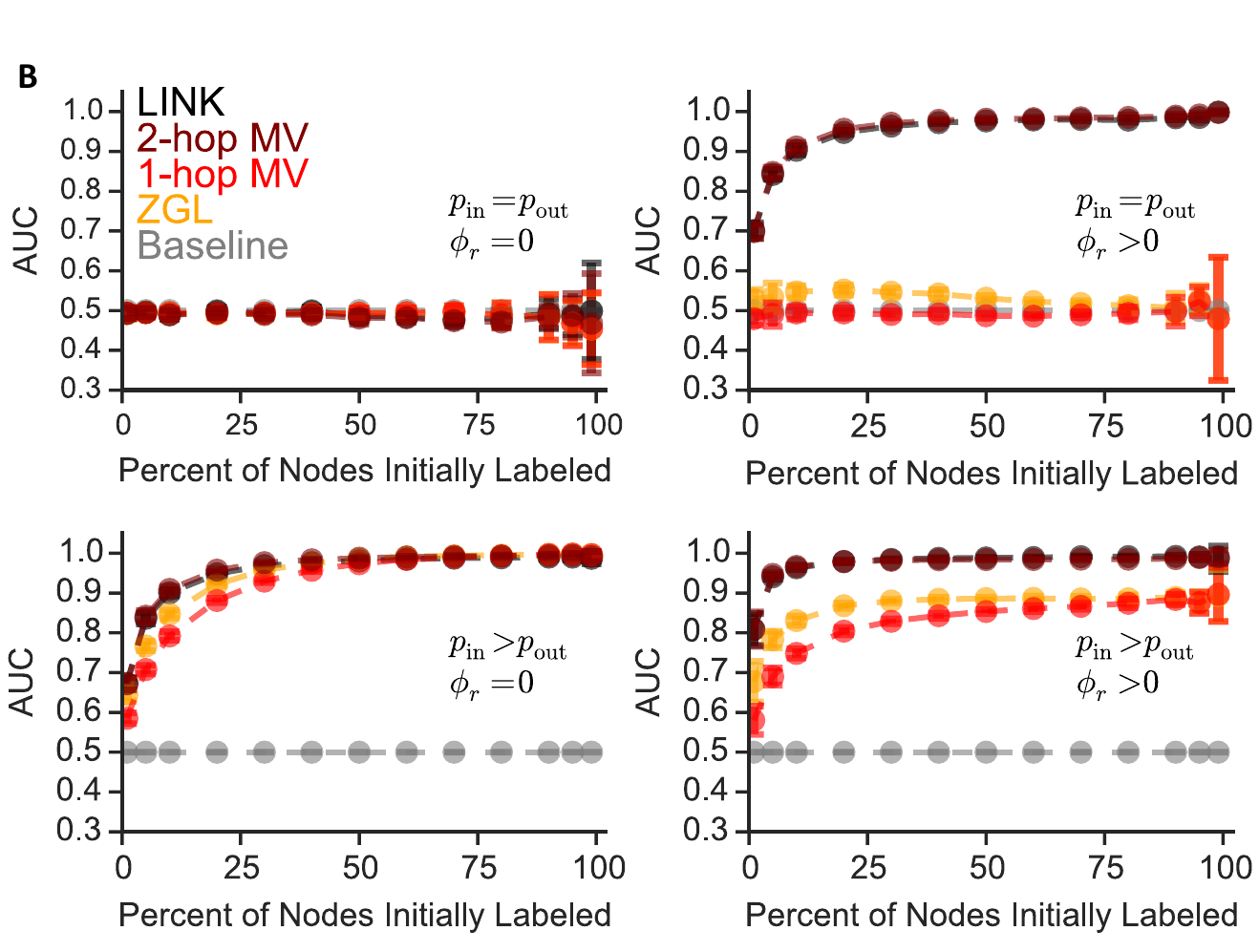}
}
\end{center}
\caption{
Four different overdispersed stochastic block models and the associated performance of 1-hop and 2-hop classifiers. (Top) Trait preference  distributions for four instances of oSBMs (filled bars) varying $p_{\text{in}}$, $p_{\text{out}}$, and $\phi_r$ parameters: no homophily and no monophily ($p_{\text{in}}=p_{\text{out}}, \phi_r = 0$), monophily but no homophily ($p_{\text{in}}=p_{\text{out}}, \phi_r > 0$), homophily but no monophily ($p_{\text{in}}>p_{\text{out}}, \phi_r = 0$), and both homophily and monophily ($p_{\text{in}}>p_{\text{out}}, \phi_r > 0$). We then compute a null distribution (solid lines) based on affinities with binomial variation (for details of null model sampling, see Methods). (Bottom) Across the same corresponding oSBM settings, we compare the relative classification performance for different inference methods and observe a clear bifurcation of performance in the case of monophily but no homophily.
}
\label{fig:oam}
\end{figure}

The oSBM allows us to validate and explore the relative performance of node inference methods on graphs with and without homophily and/or monophily. Figure \ref{fig:oam}A illustrates the distribution in gender preferences from four settings of the oSBM that vary the homophily and monophily parameters. In Figure \ref{fig:oam}B, we then compare the relative performance of 1-hop Majority Vote, ZGL, 2-hop Majority Vote, and LINK when attempting node classification on graphs from each of the four settings. We observe in the homophily-only setting ($p_{\text{in}} > p_{\text{out}}$, $\phi_r=0$) that all inference methods perform well, while in the monophily-only  setting ($p_{\text{in}} = p_{\text{out}}$, $\phi_r>0$), 1-hop MV and ZGL have no predictive power while LINK-Logistic Regression and 2-hop MV show impressive performance despite the complete lack of homophily. We conclude that the presence of monophily can be sufficient, even in the complete absence of homophily, for accurate trait inference in networks.

The overarching bias-variance framework we develop for group preferences is highly interpretable, broadly enriching the tools available for studying prediction and explanation in social systems \cite{hofman2017prediction} and helps support the continued growth of studying variation in homophily. By adapting a quasi-likelihood approach, we can simultaneously estimate both bias and overdispersion in group preferences, where the traditional homophily index and our monophily index can be interpreted as parameters within a single extra-binomial generalized linear model. This model also offers straightforward techniques for testing the statistical significance of homophily and monophily in social networks. 
 
The networks we study largely exhibit minimal gender homophily, and we attribute the success in gender prediction of the previously introduced LINK algorithm \cite{zheleva2009join} to the presence of strongly overdispersed gender preferences in these networks. We verify and generalize these empirical observations by introducing overdispersion into a stochastic block model via a multi-level approach. We use this model to demonstrate how homophily is a sufficient but not necessary condition for gender inference, and that overdispersion provides an alternative sufficient condition. This model should be of independent interest to researchers looking to create realistic models of social data that can replicate the overdispersed preferences we observe.
 
These findings provide a new perspective on social network trait classification in general and gender in particular, as well as further complicating the already difficult task of preserving privacy in social networks. The overdispersion of preferences documented in this work motivates a re-examination of 2-hop network structure in network analysis very broadly, e.g.~developing label-dependent inference methods \cite{gallagher2010leveraging} or community detection methods \cite{decelle2011asymptotic} that engage with relations among friends-of-friends, rather than only friends. Methods for studying privacy in bipartite affiliation networks \cite{kosinski2013private} should also be revisited. We ultimately believe that the overdispersion of preferences deserves study as a social structure in its own right, and encourage investigations into social correlates of preference overdispersion. While preference biases have long been the predominant focus of group structure in social networks, this work highlights the need to simultaneously give serious parallel consideration to variability.

\section*{Methods}

\subsection*{Description of Data}
We analyze populations of networks from two sources, the Facebook100 (FB100) network dataset  \cite{traud2012social} (Supplementary Note 5) and the Add Health in-school friendship nomination dataset \cite{resnick1997protecting} (Supplementary Note 9). For all networks in both datasets, we restrict the analysis to only nodes that disclose their gender, completely removing those with missing gender labels. We also restrict to nodes in the largest (weakly) connected component in order to benchmark against classification methods \cite{zhu2003semi} that assume a connected graph. The Facebook100 dataset (FB100), analyzed in the main paper, consists of online friendship networks from Facebook that was collected in September 2005 from 100 U.S. colleges, primarily consisting of college-aged individuals \cite{traud2011comparing}. We exclude Wellesley College, Smith College, and Simmons College from our analysis, which all have $>98\%$ female nodes in the original network dataset.

\subsection*{Null distribution of gender preferences}
In order to assess whether gender preferences are overdispersed in empirical networks, we compare the variance of the empirical distribution of $d_{i,\text{in}}/d_i$ across all nodes $i$ in the same class $r$ to the variance of a Binomial null distribution without overdispersion. Since the basic model assumes that $(D_{i,\text{in}} | D_i = d_i) \sim \text{Binom} (d_i, \hat{h}_r)$, we simulate draws from this distribution by repeatedly sampling from $\text{Binom} (d_i, \hat{h}_r )$ for each node $i$ to produce a distribution of samples under the null.

\subsection*{Description of cross-validation} 
We vary the percentage of initially labeled nodes by selecting a labeled sample uniformly at random \cite{macskassy2007classification}. We train our models on the $x\%$ labeled individuals (training dataset), and measure classification performance on the remaining unlabeled nodes (testing dataset), using the same train/test splits across the different inference methods. We evaluate performance for 10 different random samples of initially labeled nodes, reporting the mean weighted Area Under the Curve (AUC) for each $x\%$ of initially labeled nodes where the weights are based on the relative number of true class training labels. The vertical error bars denote the standard deviation in AUC scores across the 10 samples.  

\subsection*{Data availability}
The Facebook100 (FB100) dataset is publicly available from the Internet Archive at \\ 
\texttt{https://archive.org/details/oxford-2005-facebook-matrix} and other public repositories. The Add Health dataset can be obtained from the Carolina Population Center at the University of North Carolina by contacting \texttt{addhealth\_contracts$@$unc.edu}.

\subsection*{Code availability}
IPython notebooks are available at \url{https://github.com/kaltenburger/gender_graph_code}, documenting all results and figures.

\section*{Acknowledgements}
We thank Bailey Fosdick, Jon Kleinberg, Isabel Kloumann, Daniel Larremore, Joel Nishimura, Mason Porter, Matthew Salganik, Sam Way, and attendees of the 2016 International Conference on Computational Social Science and the 2016 SIAM Workshop on Network Science for comments. Supported in part by an National Defense Science and Engineering Graduate (NDSEG) Fellowship, the Akiko Yamazaki and Jerry Yang Engineering Fellowship, and a David Morgenthaler II Faculty Fellowship.


\bibliographystyle{plain}
\bibliography{gender_refs,supplement}

\setcounter{page}{13}
\newtheorem{makeproposition}{Proposition} 
\newtheorem{makelemma}{Lemma} 

\renewcommand{\pinfrac}{\frac{n_rp_{\text{in,r}}}{n_rp_{\text{in,r}}+n_sp_{\text{out}}}}
\newcommand{\pinfraclineindividual}{n_rp_{i,\text{in,r}}/(n_rp_{i,\text{in,r}}+n_sp_{i,\text{out}})}
\newcommand{\pinfracindividual}{\frac{n_rp_{i,\text{in,r}}}{n_rp_{i,\text{in,r}}+n_sp_{i,\text{out}}}}

\newcommand{\fix}{\marginpar{FIX}}
\newcommand{\new}{\marginpar{NEW}}
\newcommand{\abar}[2]{{f^{#1}_{#2}}}
\newcommand{\bbar}[2]{{g^{#1}_{#2}}}
\def\bra{\left\langle}
\def\ket{\right\rangle}
\def\inpathsVar{\widehat{A}} 
\def\outpathsVar{\widehat{B}} 
\def\allpathsVar{\widehat{D}} 
\def\inprobrandVar{\widehat{a}} 
\def\outprobrandVar{\widehat{b}} 
\def\covarianceVar{\mathbf{\Sigma}} 
\def\pin{p_{\text{in,r}}}
\def\pout{p_{\text{out}}}
\def\tif{\text{ if }}
\def\tand{\text{ and }}
\def\telse{\text{ else}}
\newcommand{\xhdr}[1]{{\bf #1.}}

\newpage
\section*{Supplementary Information}
\newcommand{\beginsupplement}{
        \setcounter{table}{0}
        \renewcommand{\thetable}{S\arabic{table}}
        \setcounter{figure}{0}
        \renewcommand{\thefigure}{S\arabic{figure}}
     }
\beginsupplement

\tableofcontents
\newpage

\noindent The notation is explained in the main paper, and we repeat it here for clarity. Note that we use the terminology ``nodes'' and ``individuals'' interchangeably. For notational simplicity, we will use $i \in r$ to mean the set of all nodes $i$ with attribute value $r$, $n_r$ to be the number of nodes with attribute value $r$, and $(N - n_r) = n_s$ to be the number of nodes with attribute value $s \neq r$ (where we focus primarily on a $k=2$ class set-up). The $in$-class degree $d_{i,\text{in}}$ denotes the observed number of friendships node $i$ has with individuals that also have the same attribute value $r$, and the $out$-class degree $d_{i,\text{out}}$ denotes the observed number of friendships node $i$ has with those that do not have attribute value $r$. We use capital letters ($D_{i,\text{in}},D_{i,\text{out}}$) when treating the $in$-/$out$-class degrees as random variables. Finally, we represent the probability of an $in$-class link forming as $p_{\text{in,r}}$ for nodes in class $r$ and represent the probability of $out$-class forming as $p_{\text{out}}$, where we are assuming $k=2$ classes in which case $p_{\text{out}}$ is necessarily equivalent for both classes.

\section{Distribution of $in$-class Degrees}
\label{appendix:binomial_in}

We analyze a 2-class set-up divided into attribute classes $r$ and $s$, where we give derivations for all nodes $i\in r$ and the set-up is similar for $i \in s$. For all nodes $i \in r$, node $i$'s total observed degree $d_i$ is partitioned between $in$-class degrees $d_{i,\text{in}}$ and $out$-class degrees $d_{i,\text{out}}$. We observe first that the conditional random variable $(D_{i,\text{in}} | D_i = d_i)$ is approximately binomially distributed, for all $i$ in a particular class, according to the following argument: for large populations (where $n_r$ and $N-n_r = n_s$ are large with $p_{\text{in,r}} n_{\text{r}}$ and $p_{\text{out}} n_s$ constant), then $D_{i,\text{in}}$ and $D_{i,\text{out}}$, which are binomial distributed, can be view as approximately Poisson distributed. Under this Poisson approximation, the conditional distribution $(D_{i,\text{in}}=k | D_{i,\text{in}} + D_{i,\text{in}} = d_i)$ is distributed $\text{Bin}\left(d_i,\frac{n_r p_{\text{in,r}}}{n_r p_{\text{in,r}} + n_s p_{\text{out}}}\right)$. In full formality:
\begin{eqnarray}
\Pr(D_{i,\text{in}}=k | D_i = d_i)
&=&
\Pr(D_{i,\text{in}}=k | D_{i,\text{in}} + D_{i,\text{out}} = d_i)\\
&=&
\frac{\Pr(D_{i,\text{in}} = k, D_{i,\text{out}} = d_i - k)}{\Pr(D_{i,\text{in}}+D_{i,\text{out}} = d_i)} \\
&=& 
\frac{
[ e^{- n_r p_{\text{in,r}}} \cdot \frac{(n_r p_{\text{in,r}})^k}{k!} + o(e^{-n_r}) ] 
[ e^{- n_s p_{\text{out}}} \cdot \frac{(n_s p_{\text{out}})^{d_i-k}}{(d_i-k)!} + o(e^{-n_s}) ] 
}
{
[ e^{- n_s p_{\text{out}} - n_r p_{\text{in,r}}} \cdot \frac{(n_r p_{\text{in,r}} + n_s p_{\text{out}})^{d_i}}{d_i!} + o(e^{-n_r-n_s}) ] 
} \\
&=&
\binom{d_i}{k} 
\left ( 
\frac{n_r p_{\text{in,r}}}{n_r p_{\text{in,r}} + n_s p_{\text{out}}}
\right )^k
\left ( 
\frac{n_s p_{\text{out}}}{n_r p_{\text{in,r}} + n_s p_{\text{out}}}
\right )^{d_i- k} + o(1),
\end{eqnarray}
where $o(1)$ captures an error term that is asymptotically small when $n_r$ and $n_s$ are both large. These steps allow us to identify the conditional distribution $(D_{i,\text{in}} | D_i = d_i)$ as approximately $\text{Bin}\left(d_i,\frac{n_r p_{\text{in,r}}}{n_r p_{\text{in,r}} + n_s p_{\text{out}}}\right)$. When $n_r = n_s$, this distribution reduces to simply 
$\text{Bin}\left(d_i,\frac{p_{\text{in,r}}}{p_{\text{in,r}}+p_{\text{out}}}\right)$, and when $p_{\text{in,r}}=p_{\text{out}}$, this distribution reduces simply to $\text{Bin}\left(d_i,\frac{n_r}{n_r+n_s}\right)$.

\section{Homophily Index as Intercept Term}
\label{appendix:binomial_intercept}

Here we show that the maximum likelihood estimate of the intercept term in the logistic regression model applied to the $in$- and $out$- degree counts among nodes in a particular class $r$ can be interpreted as the conventional homophily index $\hat{h}_r$. This result is derived specifically for a two-class setting.

Consider $D_{i,\text{in}} | d_i, p_{\text{in,r}}, p_{\text{out}} \sim \text{Bin}\left ( d_i,\pinfrac\right )$ for nodes $i \in r$, as derived above with $p_{\text{in,r}}$ and $p_{\text{out}}$ explicitly shown as fixed for clarity and where we define the homophily parameter $h_r=\pinfrac$. Then since the binomial distribution is a member of the exponential dispersion family and can therefore be modeled using a generalized linear model (GLM) with a logit link function, we have that \\ logit$\left(\pinfrac\right)=$log$\left(\frac{\pinfrac}{\frac{n_sp_{\text{out}}}{n_rp_{\text{in,r}} + n_sp_{\text{out}}}} \right) = $log$\left(\frac{n_rp_{\text{in,r}}}{n_sp_{\text{out}}}\right) = \beta_{0r}$ or equivalently $\pinfrac=logit^{-1}(\beta_{0r})=\frac{e^{\beta_{0r}}}{1+e^{\beta_{0r}}}$.

Given the observed degree counts for nodes with attribute value $r$ represented as $\{(d_{i,\text{in}}, d_i), i \in r \}$, which are approximately independent (but weakly dependent due to combinatorial constraints on the joint distribution of degrees), we derive the maximum likelihood estimate $\hat{\beta}_{0r}$ and show its connection with the homophily index $\hat{h}_r= \sum_{i \in r} d_{i,\text{in}} / \sum_{i \in r} d_i $. First consider the likelihood function:

\begin{eqnarray}
L(\beta_{0r}) &=& P(D_{1,\text{in}}=d_{1,\text{in}},D_{2,\text{in}}=d_{2,\text{in}},...,D_{n_r,\text{in}}=d_{n_r,\text{in}}) \\
&=& \prod_{i \in r} \binom{d_i}{d_{i,\text{in}}} \cdot \left(\pinfrac\right)^{d_{i,\text{in}}} \cdot \left(1-\pinfrac\right)^{d_{i,\text{out}}} \\
&=& \prod_{i \in r} \binom{d_i}{d_{i,\text{in}}} \cdot \left(\frac{e^{\beta_{0r}}}{1+e^{\beta_{0r}}}\right)^{d_{i,\text{in}}} \cdot \left(1-\frac{e^{\beta_{0r}}}{1+e^{\beta_{0r}}}\right)^{d_{i,\text{out}}} \\
&\propto& \left(\frac{e^{\beta_{0r}}}{1+e^{\beta_{0r}}}\right)^{\sum_{i \in r} d_{i,\text{in}}} \cdot \left(1-\frac{e^{\beta_{0r}}}{1+e^{\beta_{0r}}}\right)^{\sum_{i \in r} d_{i,\text{out}}}.
\end{eqnarray}
We transform this likelihood function to a log-likelihood function:
\begin{eqnarray}
l(\beta_{0r}) &=& log\left(\frac{e^{\beta_{0r}}}{1+e^{\beta_{0r}}}\right)\cdot\sum_{i \in r} d_{i,\text{in}} + log\left(1-\frac{e^{\beta_{0r}}}{1+e^{\beta_{0r}}}\right)\cdot\sum_{i \in r}d_{i,\text{out}}\\
&=& \beta_{0r} \cdot \sum_{i \in r} d_{i,\text{in}} - log\left(1+e^{\beta_{0r}}\right)\cdot \sum_{i \in r}d_{i,\text{in}} + log\left(1-\frac{e^{\beta_{0r}}}{1+e^{\beta_{0r}}}\right)\cdot \sum_{i \in r} d_{i,\text{out}} \\
&=& \beta_{0r} \cdot \sum_{i \in r} d_{i,\text{in}} - log\left(1+e^{\beta_{0r}}\right)\cdot \sum_{i \in r}d_{i,\text{in}} - log\left(1+e^{\beta_{0r}}\right)\cdot \sum_{i \in r} d_{i,\text{out}},
\end{eqnarray}
and from here we set $\dv{l(\beta_{0r})}{\beta_{0r}}=0$ and solve for $\hat{\beta}_{0r}$:
\begin{eqnarray}
0 &=& \sum_{i \in r} d_{i,\text{in}} - \frac{\sum_{i \in r} d_{i,\text{in}}}{1+e^{\hat{\beta}_{0r}}}\cdot e^{\hat{\beta}_{0r}} - \frac{\sum_{i \in r} d_{i,\text{out}}}{1+e^{\hat{\beta}_{0r}}}\cdot e^{\hat{\beta}_{0r}}\\
0 &=& \sum_{i \in r} d_{i,\text{in}} + \frac{e^{\hat{\beta}_{0r}}}{1+e^{\hat{\beta}_{0r}}}\cdot \left(-\sum_{i \in r} d_{i,\text{in}} -\sum_{i \in r} d_{i,\text{out}}\right) \\
\frac{e^{\hat{\beta}_{0r}}}{1+e^{\hat{\beta}_{0r}}} &=& \frac{\sum_{i \in r}d_{i,\text{in}}}{\sum_{i \in r}d_{i,\text{in}} +\sum_{i \in r} d_{i,\text{out}}} = \frac{\sum_{i \in r} d_{i,\text{in}}}{\sum_{i \in r} d_{i}}.
\end{eqnarray}
Here $\hat{\beta}_{0r}$ is the maximum likelihood estimator, and we use the superscript ``MLE'' to make this clear. Thus when using binomial logistic regression applied to the $in$-degrees $D_{i,\text{in}} | d_i, p_{\text{in,r}}, p_{\text{out}}$, we obtain that $logit^{-1}(\hat{\beta}_{0r}^{MLE})=\frac{e^{\hat{\beta}_{0r}}}{1+e^{\hat{\beta}_{0r}}} = \frac{\sum_{i \in r}d_{i,\text{in}}}{ \sum_{i \in r}d_{i}}=\hat{h}_r$, the conventional homophily index.

\section{Properties of Binomial Degree Data}
\label{appendix:dispersion}

For a realized expected degree sequence $(d_i)$ among nodes $i$ in class $r$, the conditional distribution of $in$-class degrees is (asymptotically, per Section 2): $D_{i,\text{in}} | d_i, p_{\text{in,r}}, p_{\text{out}} \sim$ Binom$\left(d_i,\pinfrac\right)$ as previously derived. In this section, we assess the unconditional expectation and variance of the $in$-class degree sequence in settings where $\pinfracindividual$ is assumed to be constant for all nodes (Model I below) and when $\pinfracindividual$ is assumed to be random (Model II below). The derivations of Model I and Model II follow those presented in Chapter 10 of \cite{garthwaite2002statistical} and are adapted to this context in terms of $in$- and $out$- class degrees.

\subsection{Without overdispersion (Model I)}
\label{appendix:no_dispersion_I}
The expectation of $D_{i,\text{in}}$ when there is no overdispersion (when $\pinfrac$ is constant for all nodes) is:
\begin{eqnarray}
\mathbb{E}\left[D_{i,\text{in}} | d_i\right] 
&=& \mathbb{E}\left[ \mathbb{E}\left[(D_{i,\text{in}} | d_i) |\pinfrac\right]\right] \\
&=& \mathbb{E}\left[ \mathbb{E}\left[d_i \cdot\pinfrac\right]\right] \\
&=& \mathbb{E}\left[ d_i \cdot \mathbb{E}\left[\pinfrac\right]\right] \\
&=& \mathbb{E}\left[ d_i \cdot\pinfrac\right] \\
&=& d_i \cdot\pinfrac.
\end{eqnarray}
The variance (again with $\pinfrac$ known) is:
\begin{eqnarray}
\Var \left[D_{i,\text{in}} | d_i\right]
&=& \mathbb{E}\left[ \Var \left[\left(D_{i,\text{in}} | d_i\right) |\pinfrac\right]\right] + \nonumber \\
&&  \Var \left[ \mathbb{E}\left[\left(D_{i,\text{in}} | d_i\right) |\pinfrac\right]\right] \\
&=& \mathbb{E}\left[d_i \cdot\pinfrac \cdot \left(1-\pinfrac\right)\right] + \nonumber \\
&&   \Var \left[d_i \cdot\pinfrac \right].
\end{eqnarray}
Considering each of these two terms, we have:
\begin{eqnarray}
&&\mathbb{E}\left[d_i \cdot\pinfrac \cdot \left(1-\pinfrac\right)\right] \\
&&= d_i \cdot \left[\mathbb{E}\left[\pinfrac\right] - \mathbb{E}\left[\left(\pinfrac\right)^2\right]\right]\\
&&= d_i \cdot \left[\mathbb{E}\left[\pinfrac\right] - \Var \left[\pinfrac\right] - \mathbb{E}\left[\pinfrac\right]^2\right] \\
&&= d_i \cdot \left[\pinfrac - 0 - \left(\pinfrac\right)^2\right]\\
&&= d_i \cdot\pinfrac \cdot \left(1 -\pinfrac\right),
\end{eqnarray}
and
\begin{eqnarray}
\Var \left[d_i \cdot\pinfrac \right]
= d_i^2 \cdot \Var \left[\pinfrac\right] 
= d_i^2 \cdot 0 = 0.
\end{eqnarray}

As a result, we obtain that $\Var \left[D_{i,\text{in}} | d_i\right] =  d_i \cdot\pinfrac \cdot \left(1 -\pinfrac\right)$.

If the expectation and variance are rewritten in terms of $h_r$, then they are: $\mathbb{E}\left[D_{i,\text{in}} | d_i\right] = d_i \cdot h_r$ and $\Var \left[D_{i,\text{in}} | d_i\right] =  d_i \cdot h_r \cdot \left(1 - h_r\right)$, respectively.

\subsection{With overdispersion (Model II)}
Following previous notational set-ups, we introduce overdispersion by allowing $\pinfracindividual$  to vary across nodes such that $\mathbb{E}\left[\pinfracindividual\right]=\mathbb{E}\left[ h_{i,r}\right]=\pinfrac=h_r$ and (for notational convenience as will be clearer later) that  $\Var \left[\pinfracindividual\right]=\Var \left[ h_{i,r}\right]$=$\phi_r \cdot\pinfrac \cdot \left(1-\pinfrac\right)=\phi_r \cdot h_r \cdot (1-h_r)$. Note that the only assumption we're making is that $\phi_r$ is constant across nodes in a given class but we are not making any distributional assumptions on $\pinfracindividual$.

Then, the unconditional expectation of $D_{i,\text{in}}$ (unconditional on $\pinfracindividual$) when there is overdispersion (when $\pinfracindividual$ is random across all nodes) is:

\begin{eqnarray}
\mathbb{E}\left[D_{i,\text{in}} | d_i\right]
&=& \mathbb{E}\left[ \mathbb{E}\left[\left(D_{i,\text{in}} | d_i\right) | \pinfracindividual\right]\right] \\
&=& \mathbb{E}\left[ \mathbb{E}\left[d_i \cdot \pinfracindividual\right]\right] \\
&=& \mathbb{E}\left[ d_i \cdot \mathbb{E}\left[\pinfracindividual\right]\right] \\
&=& \mathbb{E}\left[ d_i \cdot\pinfrac\right] \\
&=& d_i \cdot\pinfrac
\end{eqnarray}
And the unconditional variance (unconditional on $\pinfracindividual$) is:
\begin{eqnarray}
\Var \left[D_{i,\text{in}} | d_i\right]
&=& \mathbb{E}\left[ \Var \left[\left(D_{i,\text{in}} | d_i\right) | \pinfracindividual\right]\right] + \nonumber \\
&&   \Var \left[ \mathbb{E}\left[\left(D_{i,\text{in}} | d_i\right) | \pinfracindividual\right]\right] \\
&=& \mathbb{E}\left[d_i \cdot \pinfracindividual \cdot \left(1-\pinfracindividual\right)\right] + \nonumber \\
&& \Var \left[d_i \cdot \pinfracindividual \right]
\end{eqnarray}

Considering each part, we have:
\begin{eqnarray}
&& \mathbb{E}\left[d_i \cdot \pinfracindividual \cdot \left(1-\pinfracindividual\right)\right] \\
&&= d_i \cdot \left[\mathbb{E}\left[\pinfracindividual\right] - \mathbb{E}\left[\left(\pinfracindividual\right)^2\right]\right] \\
&&= d_i \cdot \left[\mathbb{E}\left[\pinfracindividual\right] - \Var \left[\pinfracindividual\right] - \mathbb{E}\left[\pinfracindividual\right]^2\right]  \\
&&= d_i \cdot \Bigg [ \pinfrac - \phi_r \cdot\pinfrac \cdot \left(1-\pinfrac\right) \\ 
&& \hspace{1.5cm} -  \left ( \pinfrac \right)^2 \Bigg ] \\
&&= d_i \cdot\pinfrac \cdot \left(1 -\pinfrac\right) \cdot \left(1-\phi_r\right), 
\end{eqnarray}
and
\begin{eqnarray}
\Var \left[d_i \cdot \pinfracindividual \right]  &&= d_i^2 \cdot \Var \left[\pinfracindividual \right] \\
&&= d_i^2 \cdot \phi_r \cdot\pinfrac \cdot \left(1-\pinfrac \right).
\end{eqnarray}

This derivation means that $\Var [D_{i,\text{in}} | d_i] =  d_i \cdot\pinfrac \cdot \left(1 -\pinfrac\right) \cdot \left(1-\phi_r\right) + d_i^2 \cdot \phi_r \cdot\pinfrac \cdot \left(1-\pinfrac\right)$, 
which simplifies to 
\begin{eqnarray}
\Var [D_{i,\text{in}} | d_i] 
&=&  d_i \cdot\pinfrac \cdot \left(1 -\pinfrac\right) \cdot \left(1-\phi_r + d_i \cdot \phi_r\right) \\ 
&=& d_i \cdot\pinfrac \cdot \left(1 -\pinfrac\right) \cdot \left(1 + \left(d_i-1\right) \cdot \phi_r\right).
\end{eqnarray}

Therefore, when $\phi_r > 0$ then the dispersion in $D_{i,\text{in}} | d_i$ is greater than what would be expected in the setting where $\phi_r=0$.
If the expectation and variance are rewritten in terms of $h_r$, then they are: $\mathbb{E}\left[D_{i,\text{in}} | d_i\right] = d_i \cdot h_r$ and $\Var [D_{i,\text{in}} | d_i] =d_i \cdot h_r \cdot \left(1 - h_r\right) \cdot \left(1 + \left(d_i-1\right) \cdot \phi_r\right)$, respectively.

\section{Algorithm for Estimating Overdispersion Parameter $\phi_r$}

This section describes the iterative procedure for estimating the overdispersion ($\phi_r$) among nodes $i$ in class $r$ by restating the procedure due to Williams \cite{williams1982extra} using our class-degree notation. The procedure initially assumes the null model without overdispersion (Model I) is true and then iteratively assesses the resulting residual variation via a goodness of fit statistic ($X^2$) based on the sum of squared residuals, allowing $\hat{\phi}_r > 0$ and then updating the estimates $\hat{\beta}_{0r}$ and $\hat{\phi}_r$ until convergence. The final $\hat{\phi}_r$ at the end of this process is the estimated overdispersion. Note that testing the goodness of fit statistic ($X^2$) relies on a predetermined significance parameter ($\alpha$), which in the main paper we test the statistical significance of overdispersion in the networks we study at the $\alpha=0.001$ significance level. For clarity on the notational differences between our work and that of Williams, we provide the following table that's explained in more detail below:

\begin{singlespace}
\begin{table}[h!]
\centering
\caption{Conversion between our notation and Williams' notation\cite{williams1982extra}.}
\label{my-label}
\begin{tabular}{|c|c|}
\hline
\textbf{Our notation} & \textbf{Williams' notation} \\ \hline
$d_i$ & $m_i$ \\ \hline
$d_{i,\text{in}}$ & $R_i$ \\ \hline
$\beta_{0r}$  
& $\lambda_i 
$ \\ \hline
$h_r = \frac{1}{1+\exp(-\beta_{0r}) } $ & $\theta = \frac{1}{1+\exp(-\lambda_i)}$ \\ \hline
\multicolumn{1}{|l|}{$\Var( h_{i,r} ) = \phi_r \cdot h_r \cdot (1-h_r)$} & $\Var (P_i) = \phi \cdot \theta_i \cdot (1-\theta_i)$ \\ \hline
\begin{tabular}[c]{@{}c@{}}
$v_i = d_i \cdot h_r \cdot (1-h_r)$ 
\end{tabular} & $v_i = m_i \cdot \theta_i \cdot (1-\theta_i)$ \\ \hline
$w_i^{-1} = 1 + \phi_r \cdot (d_i - 1)$ & $w_i^{-1} = 1 + \phi \cdot (m_i - 1)$ \\ \hline
\end{tabular}
\end{table}
\end{singlespace}

Given the observed degree data $\{(d_{i,\text{in}}, d_i), i \in r\}$ and assuming the underlying generative process is $D_{i,\text{in}} | d_i, p_{\text{in,r}}, p_{\text{out}} \sim$ Binom$\left(d_i,\pinfrac\right)$ where $d_i$ is assumed to vary across nodes, the distinguishing feature between the initial Model I (no overdispersion) and the subsequent Model II (overdispersion) is the variance: $\Var \left[D_{i,\text{in}} | d_i\right] = d_i \cdot h_r \cdot \left(1 - h_r\right) \cdot \left(1 + \left(d_i-1\right) \cdot \phi_r\right)$. For notational convenience by allowing $v_i=d_i \cdot h_r \cdot \left(1 - h_r\right)$ and $w_i^{-1}=\left(1 + \left(d_i-1\right) \cdot \phi_r\right)$, then $\Var \left[D_{i,\text{in}} | d_i\right] = v_i \cdot w_i^{-1}$ where Model I strictly enforces $\phi_r=0$ or equivalently $w_i^{-1}=w_i=1$. Meanwhile for Model II we allow $\phi_r > 0$ or equivalently $0<w_i^{-1}<1$.

The steps of Williams' iterative algorithm for jointly estimating $\hat{\beta}_{0r}$ and $\hat{\phi}_r$ are then as follows. Viewed as an iterated algorithm, the iteration is in earnest only over the variables  $\hat{\beta}_{0r}$ and $\hat{\phi}_r$ given the input data $(d_i)$ and $(d_{i,\text{in}})$, but a number of auxiliary variables (e.g.~$w_i$, $v_i$, $q$) greatly simplify the notation.

\begin{enumerate}
\item First assume there's no overdispersion ($\phi_r=0$) and test for the significance of overdispersion being present ($\phi_r > 0$) by fitting Model I assuming there are no additional explanatory variables. Then $\hat{\beta}^{MLE}_{0r} = logit(\sum_{i \in r} d_{i,\text{in}}/\sum_{i \in r} d_{i} )$. Evaluate the model fit by computing a goodness-of-fit statistic ($X^2$) or the sum of squared residuals as $X^2 =$ \\  $\sum_{i \in r} \left[(d_{i,\text{in}} - d_i \cdot \hat{h}_r)^2/(d_i \cdot \hat{h}_r \cdot (1-\hat{h}_r))\right]$, which under the null should be distributed $\chi^2_{n_r -1}$. Note as is illustrated in the main paper, we use $w_i=1$ for this initial goodness-of-fit test, a direct consequence of $\phi_r=0$ under the initial null model.  

\item Compare $X^2$ with the $\chi^2_{n_r-1}$ distribution which is valid under the null $\phi_r = 0$ being true. Then assuming the null model is true, we compute the statistical significance of $X^2$ by computing the probability that $X^2$ would be as extreme if it's assumed to be distributed $\chi^2_{n_r-1}$. If $X^2$ is significantly large as determined by the $\alpha$ significance level, then reject the null that $\phi_r = 0$ and calculate an initial estimate $\hat{\phi}^0_r$ as:
\begin{eqnarray}
\hat{h}^0_r &=& \frac{1}{1+\exp(-\hat{\beta}^{MLE}_{0r} )} \\
v^0_i &=& d_i \cdot \hat{h}^0_r \cdot (1-\hat{h}^0_r) \\
q^0 &=& \frac{1}{ \sum_{i \in r} v_i^0} \\
\hat{\phi}^{0}_r &=& \frac{X^2 - (n_r -1)
}{
\sum_{i \in r}[(d_i -1)\cdot (1-v^0_i\cdot q^0)]
}.
\end{eqnarray}

\item Update the weights $\hat{w}^{t}_i$, re-estimate $\hat{\beta}^{t}_{0r}$, and update $\hat{h}^{t}_r$ and $\hat{v}^{t}_i$: 
\begin{eqnarray}
w_i^{t+1} &=& \frac{1}{1+\hat{\phi}_r^{t} \cdot (d_i-1)}\\
\hat{\beta}^{t+1}_{0r}&=&
\frac{
\sum_{i \in r} w^{t+1}_i \cdot [ v_i^t \cdot \hat{\beta}^{t}_{0r} + d_{i,\text{in}} - d_i \cdot \hat{h}^t_r]   
}{
\sum_{i \in r} 
w_i^{t+1} v_i^t
}\\
\hat{h}_r^{t+1} &=& \frac{1}{1 + \exp (-\hat{\beta}_{0r}^{t+1}) } \\
v^{t+1}_i &=& d_i \cdot \hat{h}^{t+1}_r \cdot (1-\hat{h}^{t+1}_r).
\end{eqnarray}

\item Compute the new sum of squared residuals $X^2$:
\begin{eqnarray}
X^{2,(t+1)} = \sum_{i \in r} \frac{ 
w^{t+1}_i [d_{i,\text{in}} - d_i \cdot \hat{h}^{t+1}_r]^2
}{
v_i^{t+1}
}
\end{eqnarray}
sand if this updated value is close to the degrees of freedom $n_r - 1$, then $\hat{\phi}_r=\hat{\phi}^t_r$ is the dispersion estimate and the procedure stops. Otherwise, update 
\begin{eqnarray}
q^{t+1} &=& \frac{1}{ \sum_{i \in r} w_i^{t+1} v_i^{t+1}} \\ 
\hat{\phi}^{t+1}_r &=& 
\frac{
X^{2,(t+1)} - \sum_{i \in r} w^{t+1}_i \cdot (1-w^{t+1}_i \cdot v^{t+1}_i \cdot q^{t+1})
}{
\sum_{i \in r} w^{t+1}_i \cdot (d_i - 1) \cdot (1 - w^{t+1}_i \cdot v^{t+1}_i \cdot q^{t+1} )
}
\end{eqnarray}
and return to step 3 with $t \leftarrow t+1$.

\end{enumerate}

We use the R package \texttt{dispmod} to compute $\hat{\phi}_r$. The code snippet below assumes that the vectors \texttt{deg\_same} and \texttt{deg\_different} contains the degrees $d_{i,\text{in}}$ and $d_{i,\text{out}}$, respectively:

\footnotesize
\begin{verbatim}
    compute_monophily_phi <- function(deg_same, deg_different){
        mod <- glm(cbind(deg_same, deg_different) ~ 1, family=binomial(logit))
        mod.disp <- glm.binomial.disp(mod, maxit = 50, verbose = F)
        return(mod.disp$dispersion)
     }
\end{verbatim}
\normalsize

\section{Facebook Data Pre-processing}

\begin{figure}[h]
\begin{center}
{
	\includegraphics[width=0.75\textwidth]{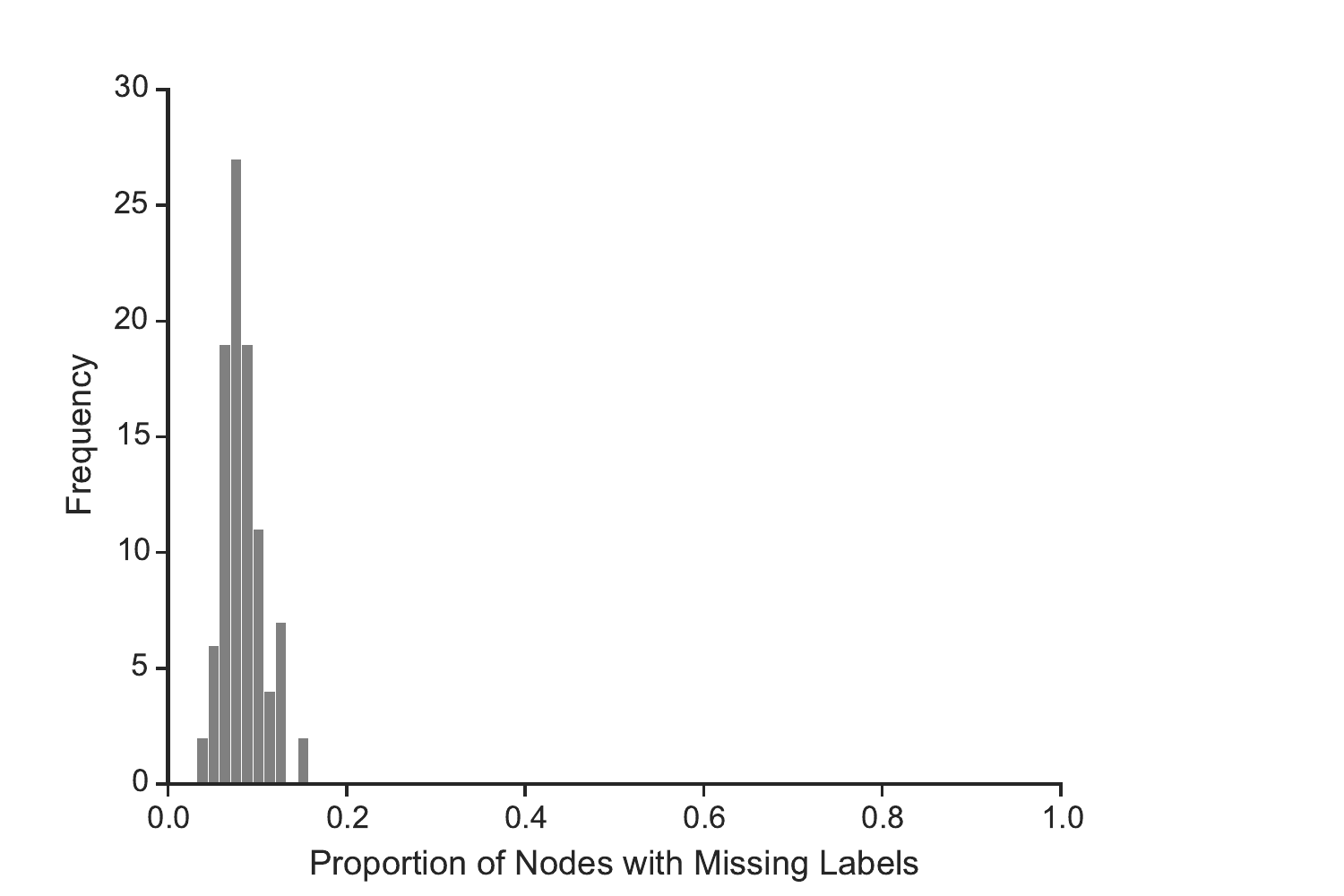}
}
\end{center}
\caption{The proportion of nodes with missing gender labels in the original network dataset across the 97 schools from the FB100 schools.}
\label{fig:fb_missing_gender}
\end{figure}

\begin{figure}[h]
\begin{center}
{
	\includegraphics[width=0.75\textwidth]{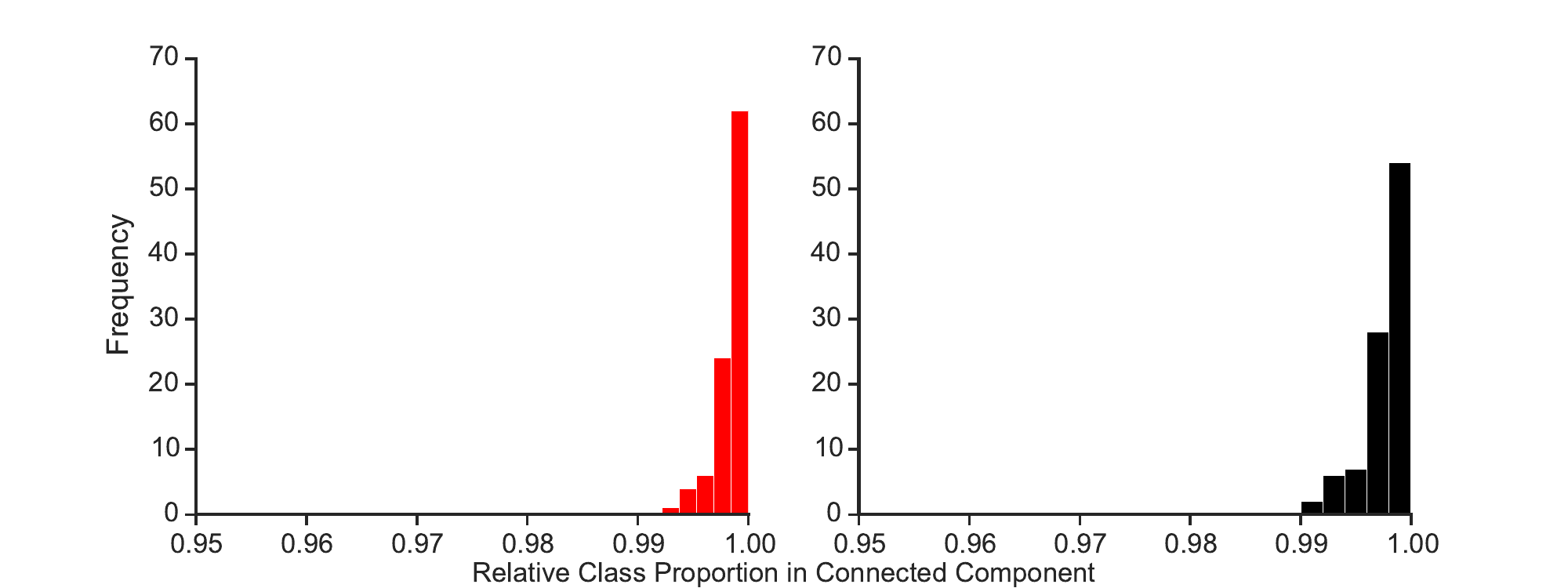}
}
\end{center}
\vspace{-0.8cm}
\caption{The relative proportion of the original nodes preserved after subsetting to the largest connected component across the 97 schools from the FB100 schools.}
\label{fig:fb_connected_component}
\end{figure}

\begin{figure}[h]
\begin{center}
{
	\includegraphics[width=0.75\textwidth]{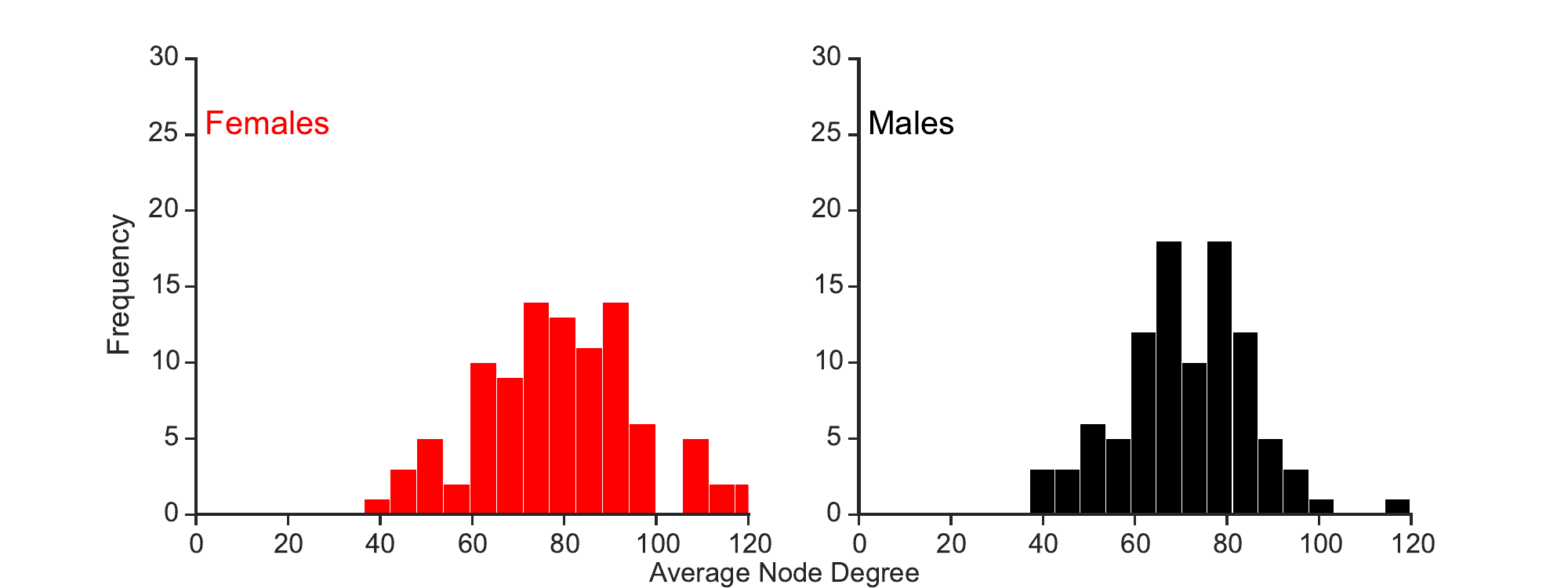}
}
\end{center}
\vspace{-0.8cm}
\caption{
The average node degree among nodes in the largest connected component across the 97 schools from the FB100 schools.}
\label{fig:fb_avg_node_degree}
\end{figure}

Figure \ref{fig:fb_missing_gender} shows that across the Facebook schools there's a very small percentage of individuals with truly unknown gender labels, always less than 16\% with an average of 8.4\%. We acknowledge that these unknown individuals can be useful for revealing the gender of others using e.g.~LINK or 2-hop Majority Vote, but given that we are not able to include them in a training/testing cross-validation set-up, we completely remove them in this work since our goal is to compare the relative performance of inference methods. Figure \ref{fig:fb_connected_component} illustrates the relative proportion of nodes in the largest connected component, and shows that these nodes comprise the majority of individuals in each graph, so subsetting is a minimal change compared to the original dataset. Figure \ref{fig:fb_avg_node_degree} illustrates that across the population of schools, males and females have comparable average degrees. The mean average degree is 71.35 for males and 79.22 for females.

\begin{figure}[h]
\begin{center}
{
	\includegraphics[width=0.5\textwidth]{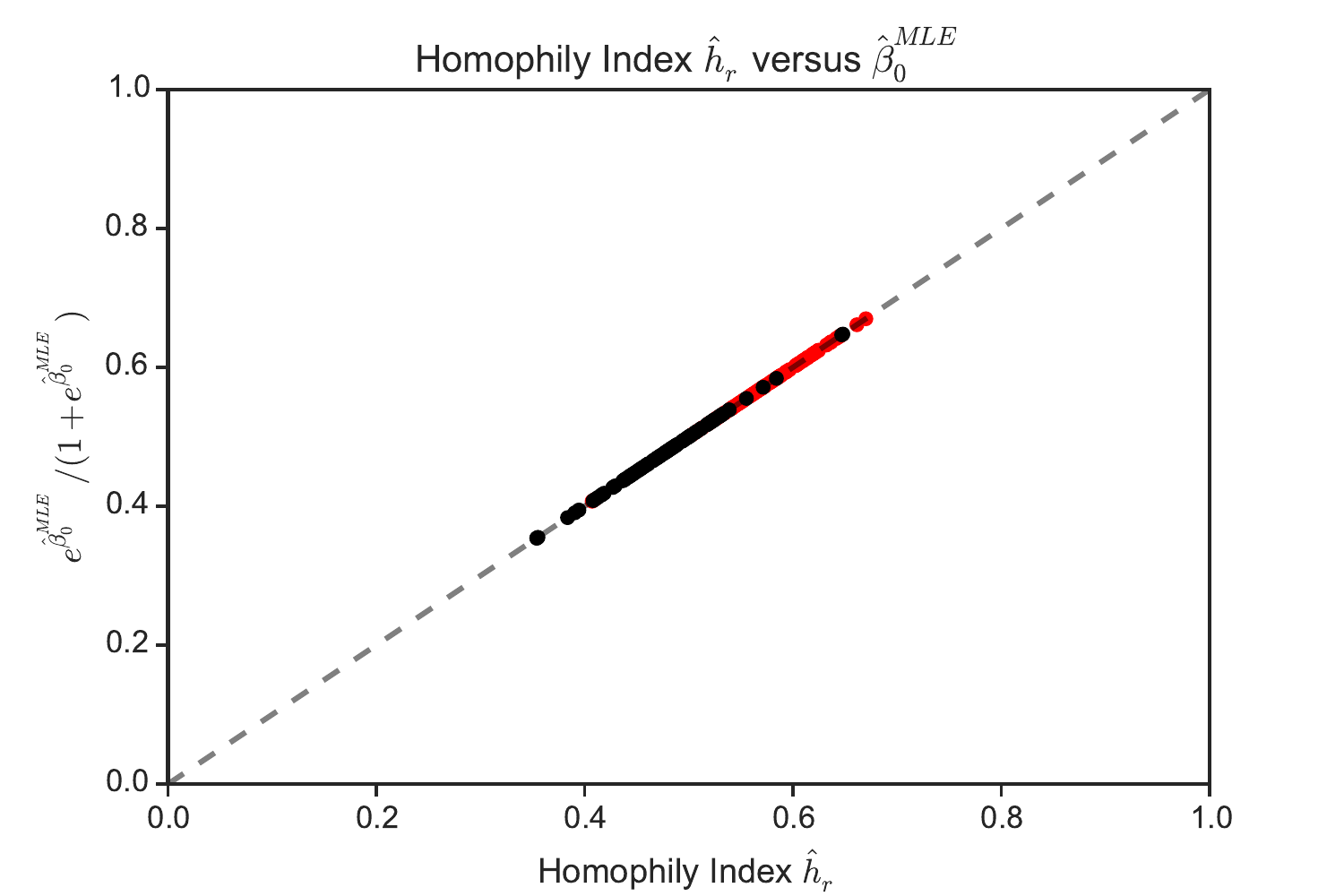}
	\includegraphics[width=0.5\textwidth]{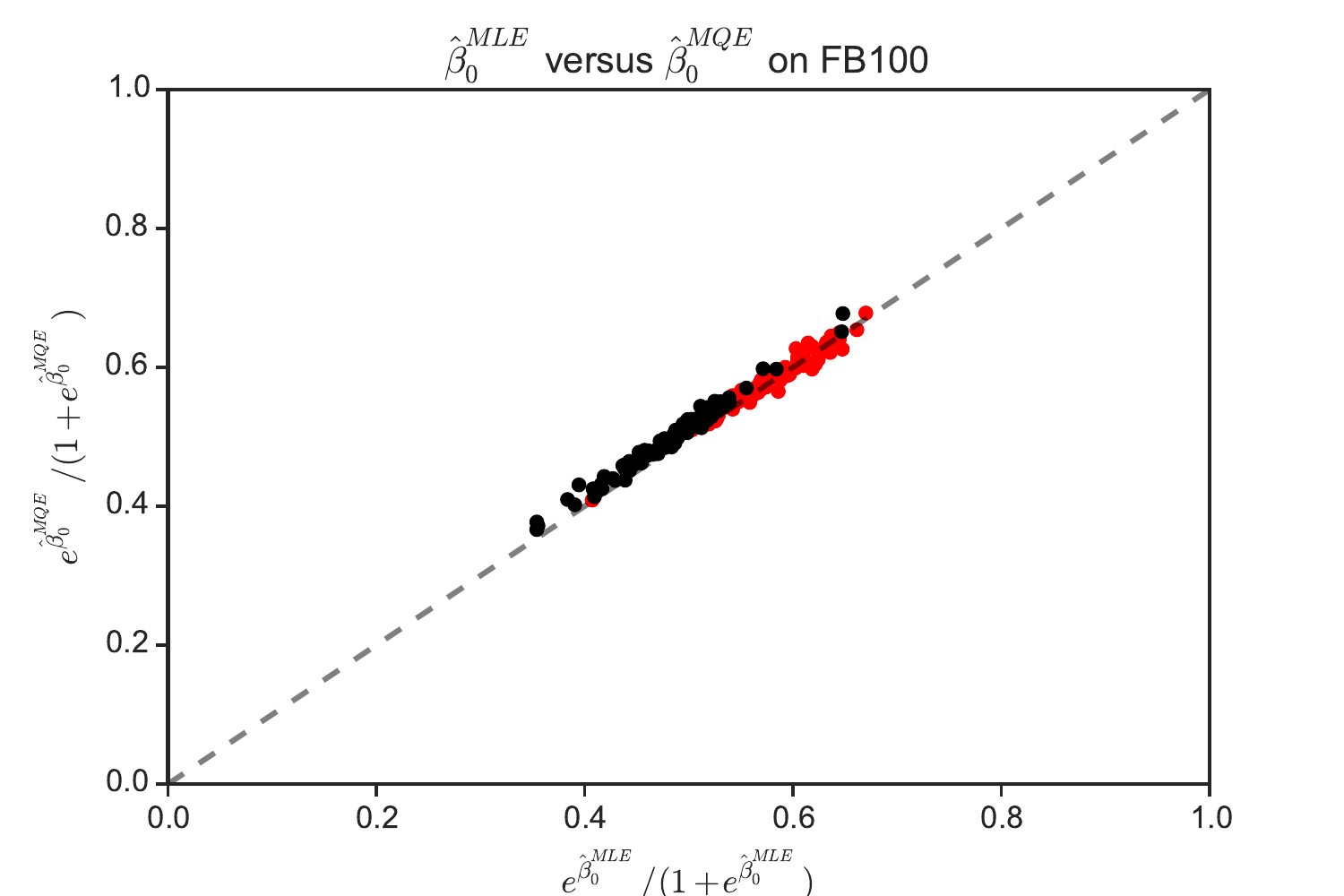}
}
\end{center}
\vspace{-0.8cm}
\caption{(Top) Confirmation that $\hat{\beta}_0^{MLE}$ and the homophily index $\hat{h}_r$ are equivalent, seen here for female (red) and male (black) classes across the 97 colleges from the FB100 networks. (Bottom) Comparison of the maximum likelihood estimate $\hat{\beta}_0^{MLE}$ versus the Williams estimate $\hat{\beta}_0^{MQE}$ obtained from maximizing the quasi-likelihood for the same data, confirming that the difference between the estimates is small in practice, as noted by Williams \cite{williams1982extra}. 
}
\label{fig:fb_b0_W_MLE}
\end{figure}

Figure \ref{fig:fb_b0_W_MLE} compares the original $\hat{\beta}^{MLE}_{0M}, \hat{\beta}^{MLE}_{0F}$ to the updated $\hat{\beta}^{MQE}_{0M}, \hat{\beta}^{MQE}_{0F}$ under the iterative Williams Method. We observe that the estimates strongly correlate, which is not surprising given that the average degree between males and females is similar and as noted in\cite{williams1982extra} that ``...the difference between the maximum likelihood estimate of $\beta$ under Model I and the maximum quasi-likelihood estimate of $\beta$ under Model II is expected to be small when the $m_i$ are of a similar magnitude.''


\clearpage

\section{Majority Vote Classification}
The aim of the 1-hop (immediate friends) and 2-hop (friends-of-friends) Majority Vote classifiers is to aggregate the known labels among an unknown node's friendship network in order to assign classification scores. Given the vector of training labels $a_1,\ldots,a_n$, where $a_i = +1$ for female training label, $a_i = -1$ for Male training label, and $a_i =NA$ is a testing label, we implement the following procedures: 

\begin{itemize}
\item For the 1-hop Majority Vote, we use the portion of the adjacency matrix corresponding to the unknown testing labels, which we'll refer to as $A_{test}$ and for a specific unknown node $u$ as $A_{u,test}=A_{test}[u,:]$. Then the classification score assigned to unknown node $u$ is based on the relative difference in the proportion of labeled male friends versus labeled female friends: 
\begin{eqnarray}
\frac{A_{u,test} \cdot \mathbb{I}[a=-1] - A_{u,test} \cdot \mathbb{I}[a=+1]}{A_{u,test} \cdot \mathbb{I}[a=-1] + A_{u,test} \cdot \mathbb{I}[a=+1]}.
\end{eqnarray} 
If node $u$ does not have any labeled neighbors meaning that $A_{u,test} \cdot \mathbb{I}[a=-1] + A_{u,test} \cdot \mathbb{I}[a=+1] = 0$, then we assign a score based on the relative proportions in the training sample: 
\begin{eqnarray}
\frac{\sum \mathbb{I}[a=-1] - \sum \mathbb{I}[a=+1]}{\sum \mathbb{I}[a=-1] + \sum \mathbb{I}[a=+1]}.
\end{eqnarray}

\item For the 2-hop Majority Vote, we implement a similar procedure as the 1-hop Majority Vote except now weights are based on $A^2_{test}$, weighted by the number of length-2 paths to labeled friends-of-friends.

\item Note that in the case of ties or when an individual has an equal number of female and male friends, then we still assign this relative class proportion since we compare relational inference methods based on their AUC.

\end{itemize}

\section{Regularization for LINK}

The LINK model introduced by Zheleva and Getoor \cite{zheleva2009join} learns a binary logistic regression classifier where the features are the entire row of the adjacency matrix among users who reveal their attribute and the outcome variable is the user's revealed attribute value. We examine these model fitting issues in this section as they pertain to binary gender inference on the datasets we study, where regularization should be given careful consideration given the large number of predictors and small number of training observations (e.g.~$p \gg n_{train}$) where $n_{train}$ denotes the size of the training sample. The method of $\ell_2$-regularized logistic regression minimizes the following cost function where $\beta$ represents the parameter vector corresponding to each of the $N$ nodes in the graph with $\beta_0$ intercept, observed gender values $y_i\in \{-1,1\}$, gain parameter $C$, and $X_i$ corresponding to the $i$th row of the adjacency matrix for user $i$:
\begin{eqnarray}
\min_{\beta_0,\beta} \frac{1}{2}\beta^T\beta + C\cdot \sum_{i \in {train}} \log(\exp(-y_i(X_i^T \beta + \beta_0))+1).
\end{eqnarray}
Here $C$ captures the inverse of the regularization strength, where for concreteness small (large) values of $C$ correspond with large (small) amounts of regularization. After exploring the sensitivity of the $C$ parameter on classification performance, we find minimal improvement from incorporating $\ell_2$-regularization. To minimize this cost function we use the implementation in Python's \texttt{scikit-learn} library.

\begin{figure}[h]
{
\begin{center}
\includegraphics[width=0.75\textwidth]{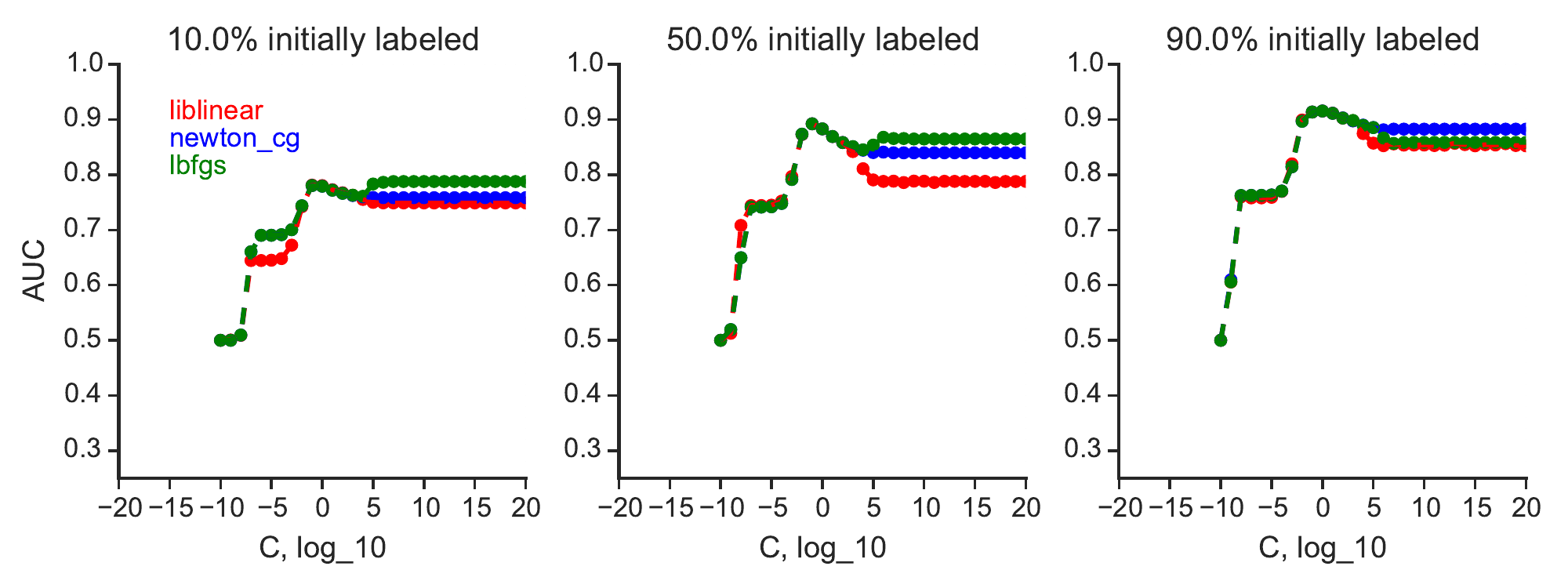}
\end{center}
}
\caption{Evaluate sensitivity to regularization parameter, $C$, on Amherst College, for 1 fold, for $\ell_2$-regularization.}
\label{fig:amherst_C_sensitivity}
\end{figure}

\begin{figure}[h]
{
\begin{center}
\includegraphics[width=0.75\textwidth]{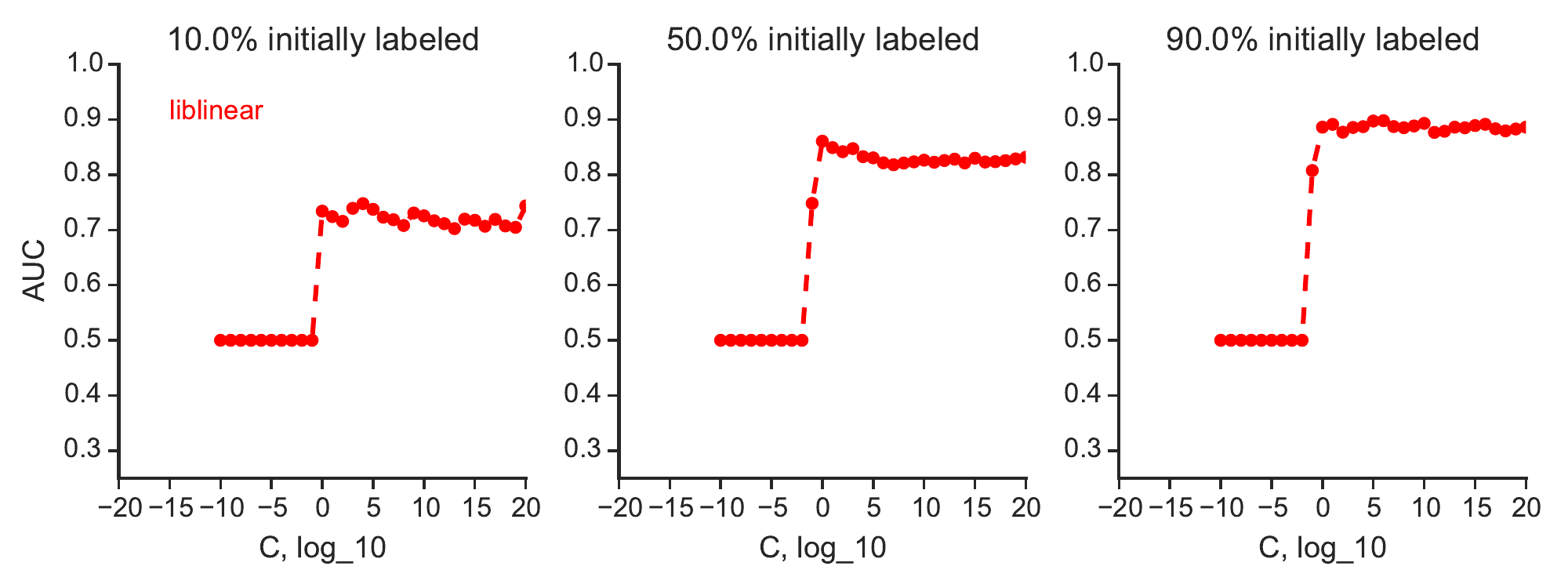}
\end{center}
}
\caption{Evaluate sensitivity to regularization parameter, $C$, on Amherst College, for 1 fold, for $\ell_1$-regularization.}
\label{fig:amherst_C_sensitivity_l1}
\end{figure}

We evaluated several different optimization methods for minimizing the regularized loss across a wide range of gains $C$; in these evaluations we focus on the Amherst College network from the FB100 dataset, the same network featured in the individual network analyses in the main paper. We evaluate \texttt{scikit-learn}'s \texttt{lbfgs}, \texttt{newton-cg}, and \texttt{liblinear} solvers, all with their default parameter settings, as illustrated in Figures \ref{fig:amherst_C_sensitivity} and \ref{fig:amherst_C_sensitivity_l1} across varying regularization gains. Note that only the \texttt{liblinear} solver is evaluated with $\ell_2$-regularization and $\ell_1$-regularization in Python, while \texttt{lbfgs} and \texttt{newton-cg} are only evaluated with $\ell_2$-regularization. These different solvers sometimes choose rather different models, as seen in the sometimes large differences in AUC. But ultimately across all three solvers we observe robust evidence that $\ell_2$-regularization does not help the AUC, and therefore choose to learn the LINK models throughout this paper (outside this section) using a very large regularization gain $C$ with the \texttt{lbfgs} solver, effectively disabling regularization.

On the subject of $\ell_1$-regularization within LINK, we briefly note that such a regularization would in a sense be trying to find a small subset of individuals to use as features for the entire graph, an insight motived by the ``subset selection'' interpretation of $\ell_1$-regularization \cite{tibshirani1996regression}. Since each of the $n$ nodes is only connected to a small fraction of the graph, there's a formal sense to which we require some $O(n)$ fraction of the nodes to have non-zero weights, not $o(n)$, contradicting the subset selection motivation. As a result, the lack of improvement from $\ell_1$-regularization is expected, and indeed this is what we see in Figure~\ref{fig:amherst_C_sensitivity_l1}.

\begin{figure}[h]
{
\begin{center}
\includegraphics[width=0.5\textwidth]{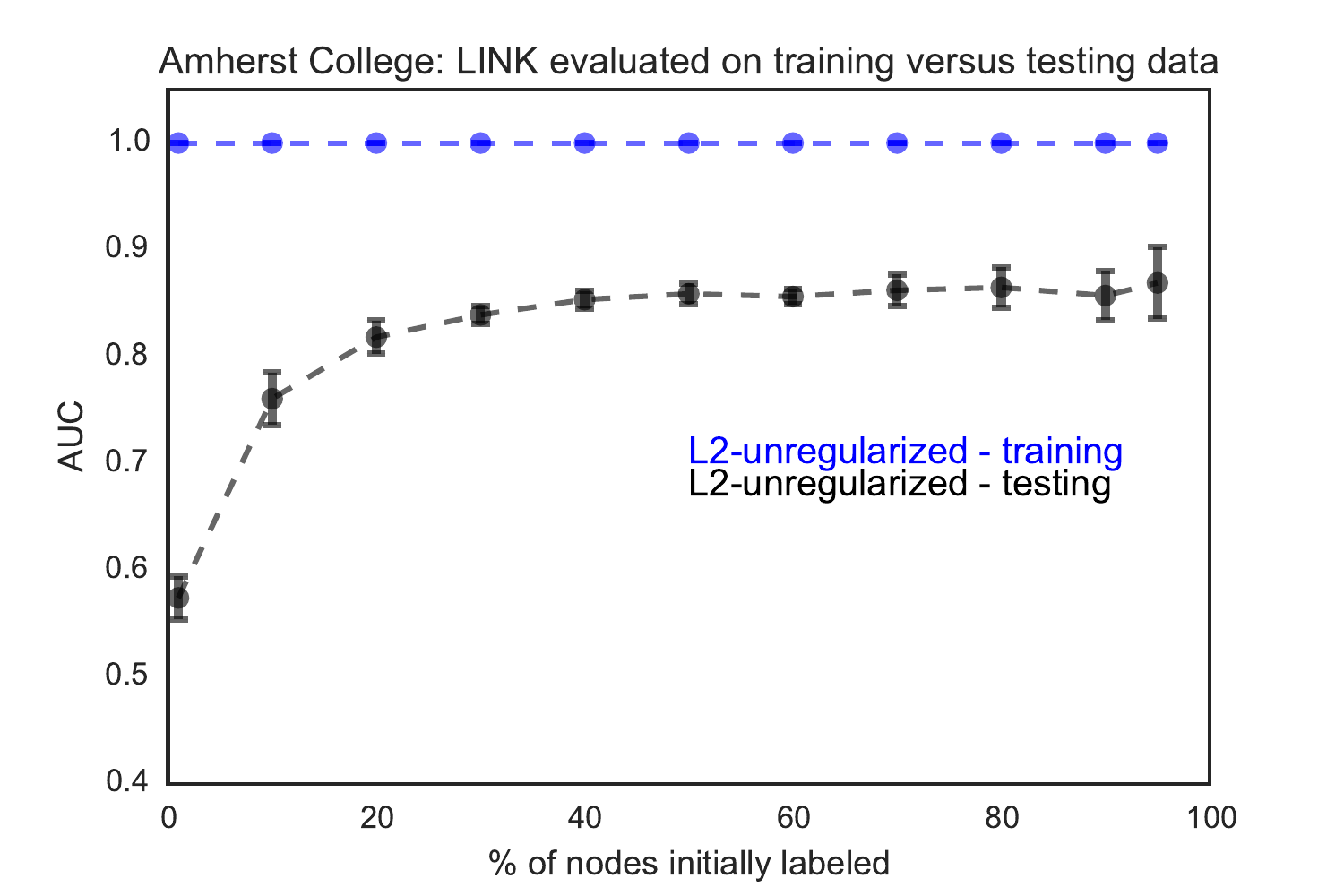}
\end{center}
}
\caption{Evaluating separability on Amherst College.}
\label{fig:amherst_separability}
\end{figure}

We observe, as has been previously noted \cite{le1992ridge}, that this unregularized model with a very large number parameters is still empirically good at distinguishing between classes. We also observe a tendency toward separability in Amherst, as seen in Figure \ref{fig:amherst_separability}. Then, as noted in \cite{rosset2004boosting}, behavior with a large gain $C$  
is similar to choosing an $\ell_2$ max-margin classifier.

\clearpage

\section{Interpretation of LINK as a 2-hop Method}

We claim that the LINK family of linear classifiers (based on linear weights applied the the columns of the adjacency matrix as feature vectors), which include LINK-Logistic Regression, LINK-SVM, and LINK-Naive Bayes, obtain their predictive power from 2-hop paths to individuals friends-of-friends. This section establishes this 2-hop connection explicitly by deriving the weights used by LINK in a Naive Bayes classifier. We demonstrate the Naive Bayes classifier reduces to a linear aggregation over a nodes friends of a nonlinear aggregation over those nodes' friends (the friends of friends of the classification subject). Note that we employ a Laplace smoothing factor (``$+1$'') to handle the case when a node does not have any male or female friends in the training sample.

Suppose that for our training data we observe labels $y_{train} \in \{M,F\}$ with the corresponding observed features $x_i \in \{0,1\}$ and random variable $X \in \{0,1\}^N$ where $N$ is the total number of nodes in the graph. This set-up represents the observed gender labels, $y_{train}$, and the observed friendships that all nodes in the network have with these training data. From this information, we construct the Naive Bayes classification rule by making the standard conditional independence assumption and studying the likelihood ratio: 
\begin{eqnarray}
LR(x) &=& \frac{P(y_{train}=F)\cdot P(X | y_{train}=F)}{P(y_{train}=M)\cdot P(X | y_{train}=M)} \\
&=& \frac{P(y_{train}=F)\cdot \prod_{i=1}^{N}P(X_i = x_i | y_{train}=F)}{P(y_{train}=M)\cdot \prod_{i=1}^{N}P(X_i = x_i | y_{train}=M)} \\
&=& \frac{P(y_{train}=F)}{P(y_{train}=M)} \nonumber \\
&&  \cdot \prod_{i:x_i = 0} \frac{P(X_i = x_i | y_{train}=F)}{P(X_i = x_i | y_{train}=M)} \cdot \prod_{i:x_i = 1} \frac{P(X_i = x_i | y_{train}=F)}{P(X_i = x_i | y_{train}=M)}. 
\end{eqnarray}

Note that in this expression we've separated the non-neighbors and neighbors of a test node by the restriction ($i:x_i = 0$ and $i:x_i = 1$), to be considered separately. 

We now have the following empirical estimates, where $n_{F,{train}}$ ($n_{M,{train}}$) denotes the number of females (males) in the training sample, $d_{i,F,{train}}$ ($d_{i,M,{train}}$) denotes node $i$'s degree with $F$ ($M$) nodes in the training sample, and we finally assume $d_{i,F,{train}} + d_{i,M,{train}} = d_{i,{train}}$. For simplicity in notation, we'll remove the ${train}$ subscript for the rest of this section. Including $+1$ Laplace smoothing we have the following standard maximum likelihood estimates for the ``parameters'' of the Naive Bayes model:  
\begin{eqnarray}
\hat{P}(y_{train}=F) &=& \frac{n_F}{n_F + n_M}\\  \hat{P}(y_{train}=M) &=& \frac{n_M}{n_F + n_M} \\
\hat{P}(X_i = 1 | y_{train}=F) &=& \frac{d_{i,F} + 1}{n_F + 2}\\ \hat{P}(X_i = 1 | y_{train}=M) &=& \frac{d_{i,M} + 1}{n_M + 2} \\
\hat{P}(X_i = 0 | y_{train}=F) &=& \frac{n_F - d_{i,F} + 1}{n_F+2}\\ \hat{P}(X_i = 0 | y_{train}=M) &=& \frac{n_M -d_{i,M} + 1}{n_M+2}.
\end{eqnarray}
Substituting these empirical estimates  into the earlier likelihood ratio, we obtain the following likelihood ratio for classifying a test node as belonging to class $F$:
\begin{eqnarray}
LR(x) &=& \frac{n_F}{n_M}  \prod_{i:x_i = 0} \frac{\left(\frac{n_F - d_{i,F}+1}{n_F+2}\right)}{\left(\frac{n_M - d_{i,M}+1}{n_M+2}\right)}  \prod_{i:x_i = 1} \frac{\left(\frac{d_{i,F}+1}{n_F+2}\right)}{\left(\frac{d_{i,M}+1}{n_M+2}\right)} \\
&=& \frac{n_F}{n_M}  \prod_{i:x_i = 0} \left(\frac{n_F - d_{i,F}+1}{n_M - d_{i,M}+1}\right) \left(\frac{n_M +2}{n_F +2}\right)  \prod_{i:x_i = 1} \left(\frac{d_{i,F}+1}{d_{i,M}+1}\right)  \left(\frac{n_M + 2}{n_F +2}\right) \\
&=& \frac{n_F}{n_M}  \left(\frac{n_M + 2}{n_F +2}\right)^N  \prod_{i:x_i = 0} \left(\frac{n_F - d_{i,F}+1}{n_M - d_{i,M}+1}\right)  \prod_{i:x_i = 1} \left(\frac{d_{i,F}+1}{d_{i,M}+1}\right) \\
&=& \frac{n_F}{n_M}  \left(\frac{n_M + 2}{n_F +2}\right)^N  \prod_{i=1}^N \left( \frac{d_{i,F}+1}{d_{i,M}+1}\right)^{x_i}  \left(\frac{n_F - d_{i,F}+1}{n_M - d_{i,M}+1} \right)^{1-x_i}.
\end{eqnarray}

Then considering the log of the likelihood-ratio, where \\ $C=\log(n_F/n_M) + N \log((n_M+2)/(n_F+2))$ is a constant:
\begin{eqnarray}
\log(LR(x)) &=& C + \log\left( \prod_{i=1}^N \left( \frac{d_{i,F}+1}{d_{i,M}+1}\right)^{x_i} \cdot \left(\frac{n_F - d_{i,F}+1}{n_M - d_{i,M}+1} \right)^{1-x_i}\right) \\
&=& C + \sum_{i=1}^N x_i \cdot \log\left(\frac{d_{i,F}+1}{d_{i,M}+1} \right) + \sum_{i=1}^N (1-x_i) \cdot \log\left(\frac{n_F - d_{i,F}+1}{n_M - d_{i,M}+1} \right) \\
&=& C +  \sum_{i=1}^N \log\left(\frac{n_F - d_{i,F}+1}{n_M - d_{i,M}+1} \right) \nonumber \\
&& + \sum_{i=1}^N x_i \cdot \left(\log\left[ \frac{d_{i,F}+1}{d_{i,M}+1} \right] - \log\left[\frac{n_F - d_{i,F}+1}{n_M - d_{i,M}+1}  \right]\right) \\
&=& C +  \sum_{i=1}^N \log\left[\frac{n_F - d_{i,F}+1}{n_M - d_{i,M}+1} \right] \nonumber \\ 
&& + \sum_{i=1}^N x_i \cdot \log\left[ \left(\frac{d_{i,F}+1}{d_{i,M}+1}\right) \cdot \left(\frac{n_M - d_{i,M}+1}{n_F - d_{i,F}+1}\right) \right].
\end{eqnarray}
If we assume that the network is sparse we have that $n_F, n_M >> d_{i,F}, d_{i,M}$ and can therefore simplify:
\begin{eqnarray}
\log(LR(x)) &\approx& \underbrace{C +  \sum_{i=1}^N \log\left[\frac{n_F}{n_M} \right]}_{C'} + \sum_{i=1}^N x_i \cdot \log\left[ \left(\frac{d_{i,F}+1}{d_{i,M}+1}\right) \cdot \left(\frac{n_M}{n_F}\right) \right] \\
&=& C' + \sum_{i=1}^N x_i \cdot \log\left[ \left(\frac{d_{i,F}+1}{d_{i,M}+1}\right) \cdot \left(\frac{n_M}{n_F}\right) \right]
\end{eqnarray}

Thus, we see that LINK-Naive Bayes in particular (and the LINK method in general) is a 2-hop method since the classification procedure relies on the degrees $d_{i,F}$ and $d_{i,M}$ of an unlabeled test node's neighbor, effectively incorporating information about the labels of nodes two hops away. In the case when $n_M=n_F$, then we can directly observe this 2-hop relation as the log-likelihood ratio $\log(LR(x))$ reduces to $C' + \sum_{i=1}^N x_i \cdot \log\left[ \left(\frac{d_{i,F}+1}{d_{i,M}+1}\right) \right]$. Ignoring the $+1$ (which comes from the use of Laplacian smoothing), the likelihood a test node is $F$ is scored based on the relative tendency of the test node's neighbors to form friendships with $F$ nodes relative to $M$ nodes. Note that the scoring is \emph{not} based on the neighbor's labels but rather on the neighbor-of-neighbor's labels in the training data.

\section{Add Health Analysis}

This section provides an analysis of gender inference on the Add Health dataset \cite{resnick1997protecting}. We evaluate homophily and monophily on the undirected and directed degree sequences, where the measures generalize cleanly to the directed setting. We follow the same data pre-processing steps as we did for the FB100, restricting the analysis to only nodes that disclose their gender and restricting to nodes in the largest (weakly) connected component.

A particularly remarkable property of the Add Health networks is that they result from a directed friendship nomination survey that limited the number of male and female friends that each person could nominate, up to five of each. While a survey under such constraints strongly limits the presence of homophily, it does not limit monophily in the in-directed network. For instance, if all females nominate person $u$, then ``having $u$ as a friend'' would be a key feature for inferring the gender of individuals who have kept that information private. We observe that it is therefore still possible to achieve high classification accuracy in the directed Add Health networks collected with constrained surveys using methods that can harness monophily, a clear demonstration that constrained surveys cannot guarantee privacy. We leave as an open question how to limit the predictive performance of directed network surveys for inferring private attributes.

The out-directed graph is described by an adjacency matrix $A_{ij}=1$ if student $i$ \emph{nominated} student $j$. The data collection was restricted such that $\sum_j A_{ij} \leq 10$, $\sum_{j \in M} A_{ij} \leq 5$, and $\sum_{j \in F} A_{ij} \leq 5$. The in-directed graph is defined by an adjacency matrix $A_{ij}=1$ if student $i$ \emph{was nominated by} student $j$ and is not restricted as $0 \leq \sum_j A_{ij} \leq N$. Due to the restriction on the out-directed degree sequences, we expect the gender preferences to be \emph{underdispersed}, which is evident in Figure \ref{fig:ah_hom}. Lastly, we drop one school (School \#27) due to the high proportion of male students (99.8\%) at that school. 

\begin{figure}[h]
\begin{center}
{
\includegraphics[width=0.5\textwidth]{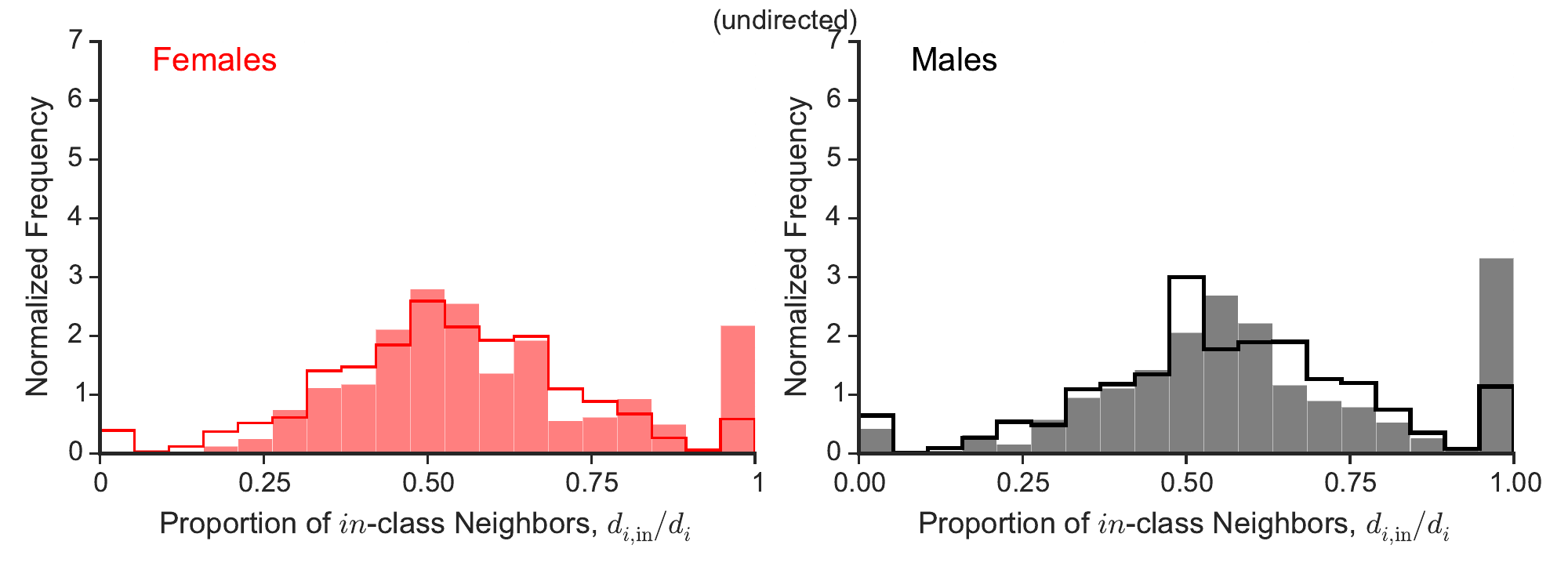}
\\

\includegraphics[width=0.5\textwidth]{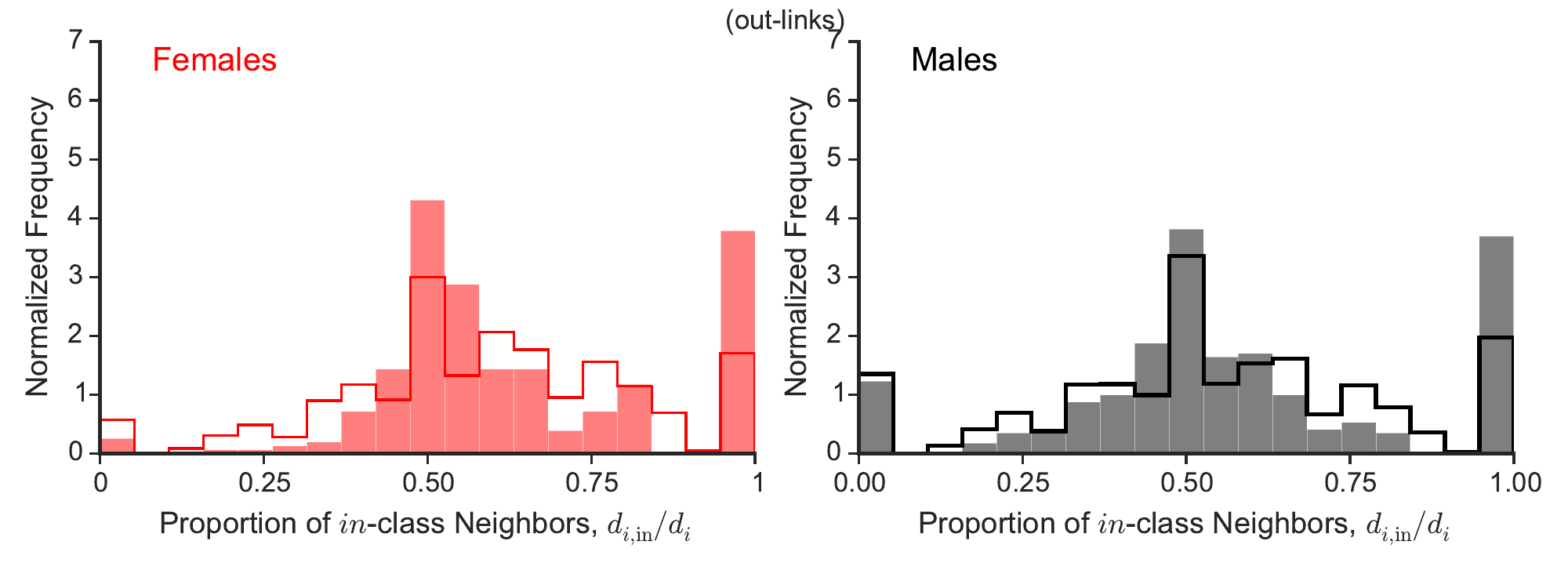}
\\

\includegraphics[width=0.5\textwidth]{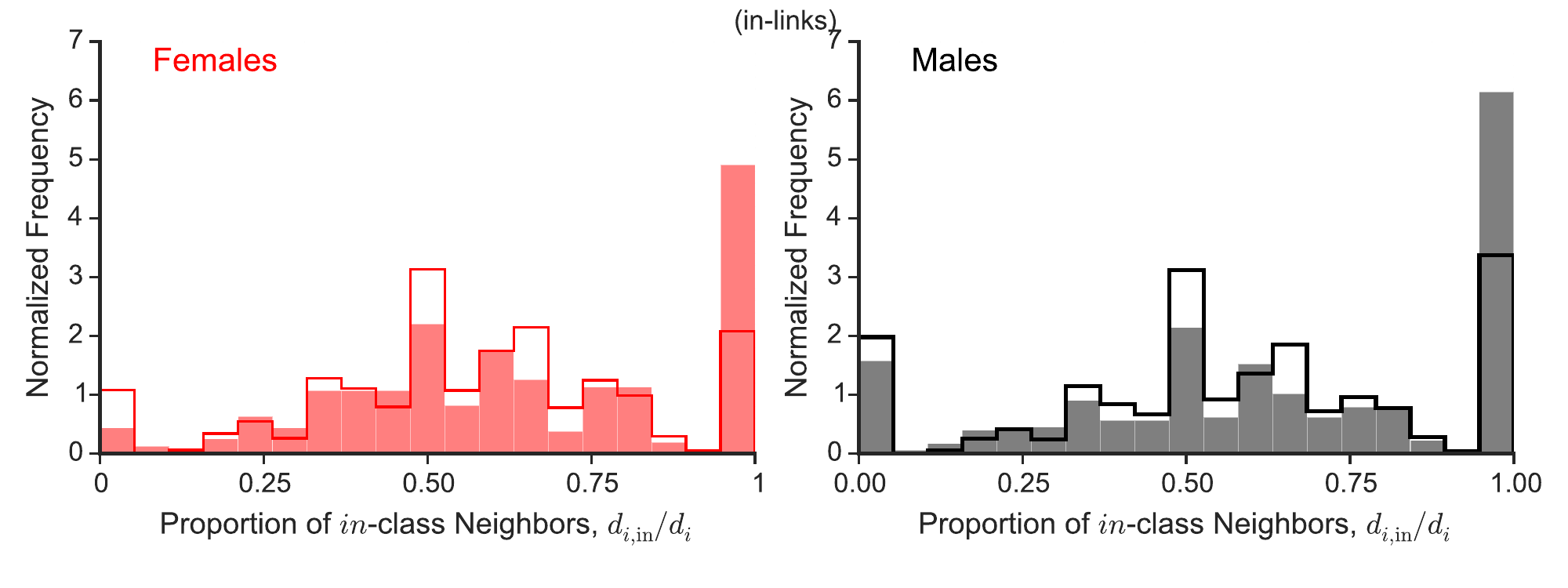}
}
\end{center}
\caption{For Add Health School \#23, we compare the variance of the empirical $in$-class preference distribution (filled bars) relative to the simulated null distribution (solid lines) for the undirected network, out-link network, and in-link network. We observe that the out-link network is underdispersed, which is in part due to the restriction on nominating male and female friends. Meanwhile, we observe the in-link network to be overdispersed which seems to be due to a larger than expected number of individuals nominating all male or all female friends.} 
\label{fig:ah_null}
\end{figure}

\begin{figure}[h]
\begin{center}
{
\includegraphics[width=0.75\textwidth]{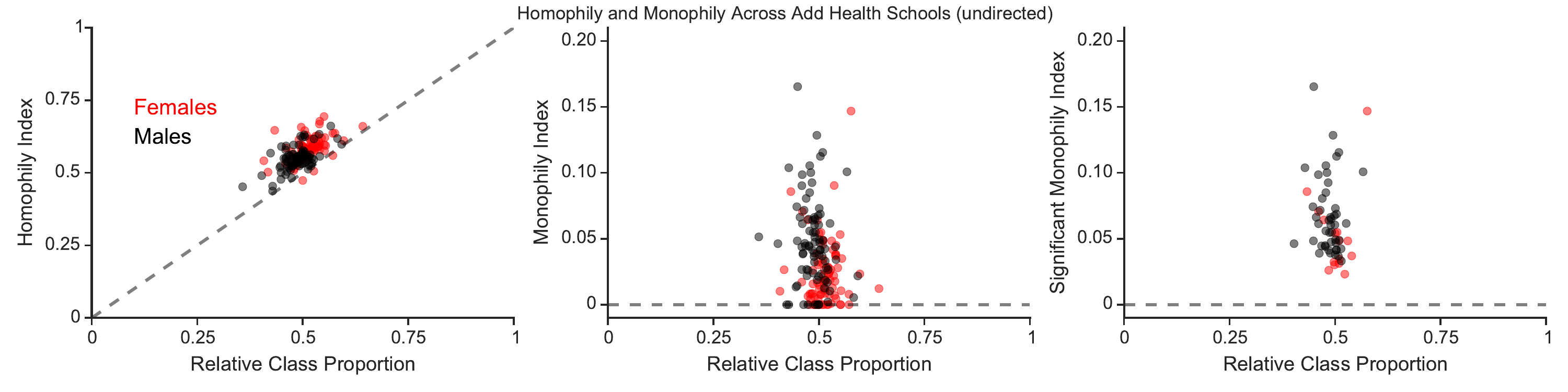}
\includegraphics[width=0.75\textwidth]{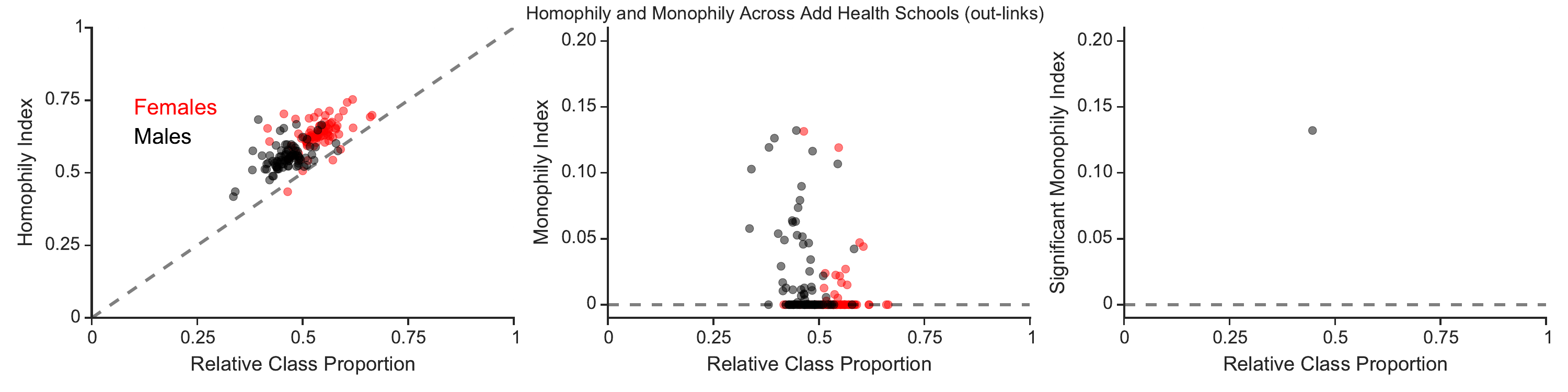}
\includegraphics[width=0.75\textwidth]{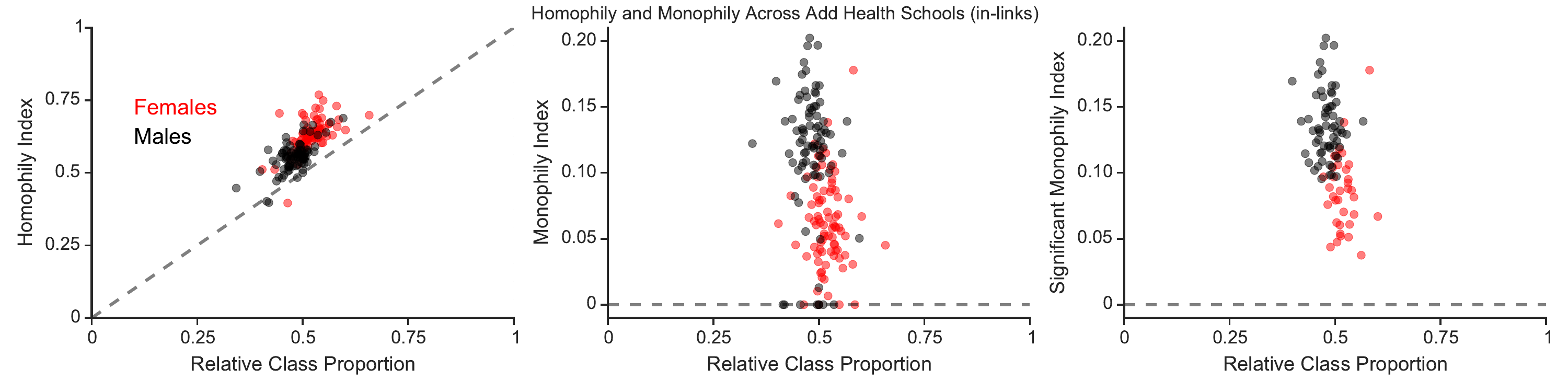}
}
\end{center}
\caption{
(Left) Homophily index $\hat{h}_r$, (Middle) monophily index $\hat{\phi}_r$, and (Right) statistically significant monophily index values at the 0.001 level across Add Health schools, in three different arrangements: (Top) undirected, (Center) out-directed, and (Bottom) in-directed.
}
\label{fig:ah_hom}
\end{figure}

Focusing on a single representative school, School \#23, we evaluate the variance of the empirical distributions for individuals to nominate same-gender friends relative to a null model as shown in Figure \ref{fig:ah_null}, which compare to Figure 1 (for the Amherst College Facebook network) in the main paper. Then evaluating the relative performance of LINK-logistic regression \cite{zheleva2009join} on the undirected versus directed degree sequences, we observe that overdispersion again drives the improved performance of LINK in the directed setting. 

\begin{figure}[h]
\begin{center}
{
	\includegraphics[width=0.5\textwidth]{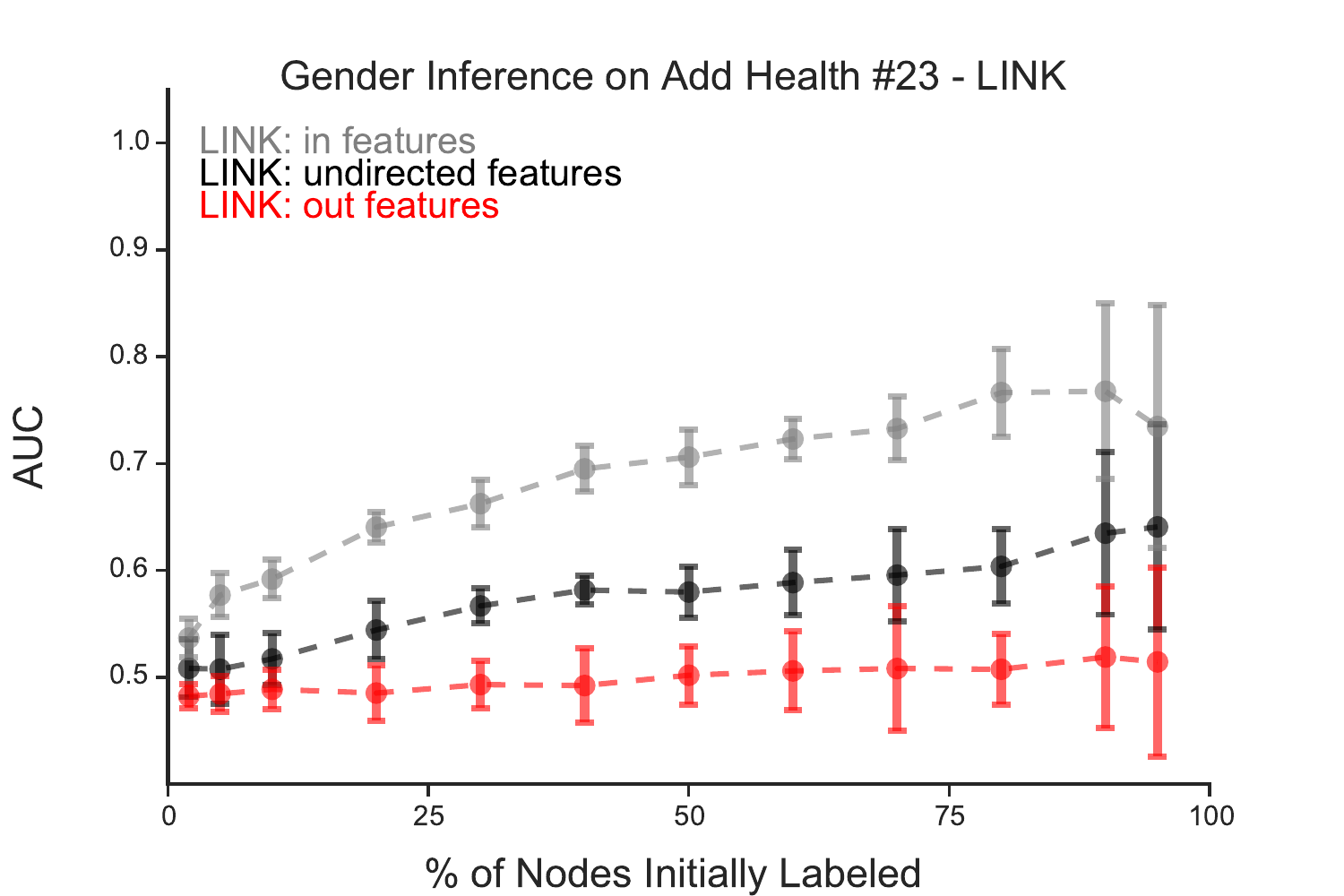}
}
\end{center}
\caption{
Gender classification on school \#23 in Add Health with $n_F$=(309,302,291) and $n_M$=(369,337,324) for the different (undirected, in-directed, and out-directed) graph versions, respectively.
}
\label{fig:ah_inference}
\end{figure}

The classification performance of LINK is driven by $in$-nominations. That is, a machine learning model based on the feature that user $i$ is nominated by all females will be more useful than a model based on the feature on who user $i$ nominates. Therefore, in Figure \ref{fig:ah_inference} we observe that the ``in features'' created based on a model fit on the out-directed adjacency matrix is the most useful in learning the relationship of correlating received nominations from particular genders. We attribute LINK's limited performance on the undirected graph as shown in Figure \ref{fig:ah_inference} to the underdispersion inherent in the Add Health data collection process.

\clearpage

\section{Sampling graphs from the Overdispersed Stochastic Block Model (oSBM)}

This section introduces an algorithm for sampling graphs from an overdispersed stochastic block model on $k$ blocks assuming a latent Beta distribution on friendship preferences and adopting the Beta parameterization in \cite{prentice1986binary}. The algorithm takes as input parameters to control the block structure, $p_{\text{in}}$ and $p_{\text{out}}$, as well as parameters to control the dispersion, $\phi^*_{\text{in}}$ and $\phi^*_{\text{out}}$. Note we assume the same parameter value for $p_{in,r}$ across all attribute classes $r$, so denote this by $p_{in}$ instead of $p_{in,r}$ for a given class $r$. In settings where one wants to preserve a given overall average node degree $\bar{d} = \frac{1}{N}\cdot \sum_{i=1}^{N} d_i$, we give the following parameterization for $p_{\text{in}}$ and $p_{\text{out}}$ that relies on a block structure parameter $\lambda \ge 1$ and assumes $k>1$ blocks:
\begin{eqnarray}
\frac{\lambda}{N}=p_{\text{in}} /\ \bar{d} &\implies& p_{\text{in}}=\lambda\cdot \frac{\bar{d}}{N} \\
\bar{d} =\frac{1}{N} \cdot \sum_{i=1}^{k} n_i \cdot (p_{\text{in}}\cdot n_i + p_{\text{out}}\cdot \sum_{j=1,j\neq i}^{k} n_j)) &\implies& p_{\text{out}}=\frac{\bar{d}\cdot N - p_{\text{in}} \cdot \sum_{i=1}^{k} {n_i}^2 }{\sum_{i=1}^{k} n_i \cdot (\sum_{j=1,j\neq i}^{k} n_j)}.
\end{eqnarray}

\linespread{1}
\begin{algorithm}
\caption{Sample from Overdispersed Block Model}
\label{oSBM}
\begin{algorithmic}[1]
\Procedure{Sample from Overdispersed Block Model}{}\\
Input: Given $k\geq 2$ mutually exclusive attribute class labels \{$a_1$,...,$a_k$\} for unique class blocks of size $n_1$,...,$n_k$ nodes, respectively, let $N$=$\sum_{j=1}^{k} n_j$. Also given: block structure $p_{\text{in}}, p_{\text{out}}$, dispersion parameters $\phi^*_{\text{in}}, \phi^*_{\text{out}}$.

\State Model affinity probabilities for $in$- and $out$-class degree distributions using latent Beta distribution \cite{prentice1986binary} with parameters $0<p_{\text{in}}<1$ and $0<p_{\text{out}}<1$: 
\State{Create $in$-class parameters:}
\State \indent set $\alpha_{\text{in}}=p_{\text{in}}\cdot(\frac{1}{\phi^*_{\text{in}}}) \cdot (1-\phi^*_{\text{in}})$
\State \indent set $\beta_{\text{in}}=(1-p_{\text{in}}) \cdot \left(\frac{1}{\phi^*_{\text{in}}}\right)\cdot\left(1-\phi^*_{\text{in}}\right)$
\State{Create $out$-class parameters:}
\State \indent set $\alpha_{\text{out}}=p_{\text{out}}\cdot\left(\frac{1}{\phi^*_{\text{out}}}\right) \cdot \left(1-\phi^*_{\text{out}}\right)$
\State \indent set $\beta_{\text{out}}=\left(1-p_{\text{out}}\right) \cdot \left(\frac{1}{\phi^*_{\text{out}}}\right)\cdot\left(1-\phi^*_{\text{out}}\right)$

\For{each attribute class $\{1,...,k\}$}
	\For{each node $i$ in specific class $r$ }
		\State $p_{i,\text{in}} \sim \text{Beta}(\alpha_{\text{in}},\beta_{\text{in}})$
	\State $d_{i,\text{in}} = p_{i,\text{in}} \cdot n_r$ 
	\item[]
	\State $ p_{i,\text{out}} \sim \text{Beta}(\alpha_{\text{out}},\beta_{\text{out}})$
	\State $d_{i,\text{out}} = p_{i,\text{out}} \cdot \sum_{j=1,j\neq r}^{k} n_j = p_{i,\text{out}} \cdot (N-n_r)$ 
	\EndFor
\EndFor

\For{each affiliation pair $a_i, a_j $ s.t. $a_i \leq a_j$} 
\If{$a_i = a_j = r $}
	\State Create Chung-Lu graph with expected degree sequence $(d_{i,\text{in}})$ $\forall$ $i \in r$:
	\For{each node pair $i, j \in r$, $i \le j$:} 
		\State $A_{ij}|p_{i,\text{in}},p_{j,\text{in}} \sim \text{Bern}\left(\frac{d_{i,\text{in}}d_{j,\text{in}}}{n_r^2\cdot p_{\text{in}}}\right)$ 
		\State Set $A_{ji} := A_{ij}$ 
	\EndFor
\EndIf
\If{$a_i = r \neq a_j = s$}
	\State Create bipartite Chung-Lu graph (adopting \cite{larremore2014efficiently}) with expected degree sequence $(d_{i,\text{out}})$ and $(d_{j,\text{out}})$ $\forall i\in r$, $\forall j \in s$:
	\For{each node pair $i \in r$, $j \in s$:} 
		\State $A_{ij}|p_{i,\text{out}},p_{j,\text{out}}  \sim \text{Bern} \left (
		\frac{d_{i,\text{out}}d_{j,\text{out}}}{(N-n_r) \cdot (N-n_s) \cdot p_{\text{out}}}\right)$,
		\State Set $A_{ji} := A_{ij}$
	\EndFor
\EndIf
\EndFor
\State Output: Symmetric adjacency matrix $A$.
\EndProcedure
\end{algorithmic}
\end{algorithm}
\linespread{2}

\newpage

\subsection{Properties based on oSBM parameterization.}

We assume an underlying model where nodes have an individual affinity for $in$- and $out$-class friends where the mean $in$-class affinity is $p_{\text{in}}$ and the mean $out$-class affinity is $p_{\text{out}}$. For all nodes $i$ in class $r$, we draw $in$- and $out$-class affinity probabilities from a Beta distribution with the following parameterization to preserve the mean $in$- and $out$- affinities and the given dispersion parameters $\phi_{\text{in}}$ and $\phi_{\text{out}}$:\\
\\
$
p_{i,\text{in}} \sim \text{Beta}(\alpha_{\text{in}}, \beta_{\text{in}})
$
\begin{eqnarray}
\text{where }\alpha_{\text{in}}&=&p_{\text{in}}\cdot\left(\frac{1}{\phi_{\text{in}}}\right) \cdot \left(1-\phi_{\text{in}}\right)\\
\beta_{\text{in}}&=&(1-p_{\text{in}}) \cdot \left(\frac{1}{\phi_{\text{in}}}\right)\cdot\left(1-\phi_{\text{in}}\right) = \left(\frac{1}{\phi_{\text{in}}}\right)\cdot\left(1-\phi_{\text{in}}\right) - \alpha_{\text{in}},
\end{eqnarray}
$
p_{i,\text{out}} \sim \text{Beta}(\alpha_{\text{out}}, \beta_{\text{out}})
$
\begin{eqnarray}
\text{where }\alpha_{\text{out}}&=&p_{\text{out}}\cdot\left(\frac{1}{\phi_{\text{out}}}\right) \cdot \left(1-\phi_{\text{out}}\right)\\
\beta_{\text{out}}&=&\left(1-p_{\text{out}}\right) \cdot \left(\frac{1}{\phi_{\text{out}}}\right)\cdot\left(1-\phi_{\text{out}}\right)
= \left(\frac{1}{\phi_{\text{out}}}\right)\cdot\left(1-\phi_{\text{out}}\right) - \alpha_{\text{out}}.
\end{eqnarray}

\subsection{Confirm mean attribute affinity is preserved.}
For $in$-class attribute affinity probabilities, we confirm that the above parameterization of the latent Beta distribution is such that $\mathbb E[p_{i,\text{in}}] = p_{\text{in}}$:
\begin{align}
\mathbb E [p_{i,\text{in}}] = \frac{\alpha_{\text{in}}}{\alpha_{\text{in}}+\beta_{\text{in}}} = \frac{p_{\text{in}}\cdot\left(\frac{1}{\phi_{\text{in}}}\right) \cdot \left(1-\phi_{\text{in}}\right)}{\left(\frac{1}{\phi_{\text{in}}}\right)\cdot\left(1-\phi_{\text{in}}\right)} = p_{\text{in}}.
\end{align}
A similar check shows that the $out$-class affinity probability is such that $\mathbb E[p_{i,\text{out}}] = p_{\text{out}}$.

\subsection{Confirm variance of attribute affinity is preserved.}

For $in$-class affinity probabilities, we confirm that the above parameterization is such that $\phi_{\text{in}}$ introduces extra binomial variation such that $\Var [ p_{i,\text{in}} ] = \phi_{\text{in}} \cdot p_{\text{in}} \cdot (1-p_{\text{in}})$:

\begin{align}
\Var [p_{i,\text{in}}]
&=\frac{\alpha_{\text{in}}\beta_{\text{in}}}
{(\alpha_{\text{in}}+\beta_{\text{in}})^2(\alpha_{\text{in}}+\beta_{\text{in}}+1)}
\\
&= \frac{\alpha_{\text{in}}}
{\alpha_{\text{in}}+\beta_{\text{in}}}
\frac{\beta_{\text{in}}}
{\alpha_{\text{in}}+\beta_{\text{in}}}
\frac{1}{\alpha_{\text{in}} + \beta_{\text{in}}+1}
\\
&=p_{\text{in}} \cdot (1-p_{\text{in}}) \cdot \frac{1}{\frac{1}{\phi_{\text{in}}}\cdot(1-\phi_{\text{in}})+1} 
\\
&=p_{\text{in}} \cdot (1-p_{\text{in}}) \cdot \frac{1}{\frac{1}{\phi_{\text{in}}}-1+1}
\\
&=p_{\text{in}} \cdot (1-p_{\text{in}}) \cdot \phi_{\text{in}}.
\end{align}

\noindent A similar check for the out-community affinity probability shows that $\Var [ p_{i,\text{out}} ] = \phi_{\text{out}} \cdot p_{\text{out}} \cdot (1-p_{\text{out}})$.

\subsection{Confirm that expected degrees are approximately preserved.}
We confirm the oSBM parameterization for edge probabilities, outlined in Algorithm \ref{oSBM} such that $P(A_{ij}=1| d_{i,\text{in}},d_{j,\text{in}})$ for $in$-class edges or $P(A_{ij}=1| d_{i,\text{out}},d_{j,\text{out}})$ for $out$-class edges, approximately preserves $(d_{i,\text{in}})$ and $(d_{i,\text{out}})$ as the class-specific expected degrees, and will therefore also approximately preserve $(d_{i})$ as the overall expected degrees. 

Let node $i$ be in class $r$, and we want to show that $\mathbb{E}\left[\sum_{j \in r}A_{ij} | p_{i,\text{in}}, p_{j,\text{in}}\right] = d_{i,\text{in}}$. We have:
\begin{eqnarray}
\mathbb{E}\left[\sum_{j \in r}A_{ij} | p_{i,\text{in}}, p_{j,\text{in}}\right] 
&=& \sum_{j \in r} \mathbb{E}[A_{ij}| p_{i,\text{in}}, p_{j,\text{in}}] \\
&=& \sum_{j \in r} \frac{d_{i,\text{in}} \cdot d_{j,\text{in}}}{n_{r}^2 \cdot p_{\text{in}}} \\
&=& \frac{d_{i,\text{in}}}{n_r \cdot p_{\text{in}}} \sum_{j \in r} \frac{p_{j,\text{in}} \cdot n_r}{n_{r}} \\
&=& \frac{d_{i,\text{in}}}{n_r \cdot p_{\text{in}}} \sum_{j \in r} p_{j,\text{in}} \\
&\approx & \frac{d_{i,\text{in}}}{n_r \cdot p_{\text{in}}} \left (n_r \cdot p_{\text{in}} \right ) \\
&=& d_{i,\text{in}}.
\end{eqnarray}

Let node $i$ be in class $r$ and $\mathbb{E}\left[\sum_{j \in s\neq r}A_{ij} | p_{i,\text{out}}, p_{j,\text{out}}\right] = d_{i,\text{out}}$, then we have the following:

\begin{eqnarray}
\mathbb{E}\left[\sum_{j \in s\neq r}A_{ij} | p_{i,\text{out}}, p_{j,\text{out}}\right] &=& \sum_{j \in s\neq r} \mathbb{E}[A_{ij}|p_{i,\text{out}}, p_{j,\text{out}}] \\
&=& \sum_{j \in s\neq r} \frac{d_{i,\text{out}}\cdot d_{j,\text{out}}}{(N-n_r) \cdot (N-n_s) \cdot p_{\text{out}}} \\
&=& \sum_{j \in s\neq r} \frac{p_{i,\text{out}} \cdot (N-n_r) \cdot p_{j,\text{out}} \cdot (N-n_s)}{(N-n_r) \cdot (N-n_s) \cdot p_{\text{out}}}  \\
&=& \frac{p_{i,\text{out}}}{p_{\text{out}}} \cdot \sum_{j \in s\neq r} p_{j,\text{out}} \\
&\approx & \frac{p_{i,\text{out}}}{p_{\text{out}}} \cdot (N-n_r) \cdot p_{\text{out}} \\
&=& p_{i,\text{out}} \cdot (N-n_r)\\
&=& d_{i,\text{out}}.
\end{eqnarray}


\end{document}